\documentclass[preprint,aps,twocolumn,showpacs,prb,10pt]{revtex4}
\usepackage{amsfonts}
\usepackage{graphicx}
\usepackage{amsmath}
\usepackage{natbib}
\usepackage{epsfig}
\usepackage{dcolumn}

\begin{document}

\title{Dynamics of a trapped domain wall in a current perpendicular
to the plane spin valve nano-structure}

\author{A. Rebei}

\email{arebei@mailaps.org}

\author{O. Mryasov}

\email{onmryasov@zoominternet.net}

\affiliation{Seagate Research Center, Pittsburgh, Pennsylvania
15222,USA}


\date{May 5, 2006}

\begin{abstract}
A study of transverse tail-to-tail magnetic domain walls (DW) 
in novel  current perpendicular
to the plane (CPP) spin 
valves (SV) of various dimensions is presented.  For  films 
with  dimensions
larger than the DW width, we find that DW motion can give rise to
a substantial low frequency noise.  For dimensions comparable to
the DW width, we show that  the DW can be controlled
by an external field or by a spin momentum torque as opposed to
the case of  CPP-SV
with uniform magnetization.  It is shown that
in a single domain biased CPP-SV,
 the spin torque can give rise to 1/f-type noise. The
  dipolar
field, the spin torque and the Oersted field are all 
accounted for in this work.  Our
proposed SV requires low current densities to move DW
and can simulate devices for
logical operation or magnetic sensing without having to
switch the magnetization in the free layer.
\end{abstract}

\pacs{xxx}

\maketitle

\section {Introduction}
\label{sec:intro}

The study of domain walls (DW) magnetic  (static and dynamic)
\cite{bruno,molyneux,hertel,saitoh} and transport
\cite{berger2,zhang,simanek,rippard} properties have attracted
much attention recently due to their  relevance for
magneto-electronic nano-device 
applications. In the static case, McMichael and Donahue have 
shown that in magnetic stripes head-to-head DW can be in vortex or transverse 
 shape \cite{mcmicheal}. Klaui \textit{et al.} have 
observed these DW structures in magnetic thin films and 
rings \cite{klaui}. \ In this work, we study the
 dynamics of DW with magnetization
mostly restricted to the plane of a nanometer-size thin film.
\   DW motion can  be a 
source of low frequency noise in
magnetic systems and thus hinder any potential applications. Their
high frequency behavior can however be tuned by choosing suitable
boundary conditions or geometries. Constrained DW oscillations can
be high in frequency  and hence there is a potential for their use
in nano-electronics as resonators \cite{rippard}.

At least two modes of oscillations have been  studied in the
literature; a Doring-type oscillation  and a Winter mode
oscillation \cite{doring,welch,winter,thiele}.
The Doring mode is associated with translations of the center of
'mass' of the DW in an infinite system, while the Winter modes are
non-zero energy modes that, in addition to rigid translation, 
correspond
to propagations along the DW.
These modes are  found by solving the time-dependent Landau-Lifshitz (LL)
equations \cite{landau,malozemoff}.

 In the following, we focus only on dynamical 
properties of
  transverse {\it{tail-to-tail}} DW trapped  in stripes
 with dimensions
comparable to their width. This case is
proved to be the most  interesting due to the unusual magnetization
dynamics as opposed to the conventional case of DW with widths being
much smaller than the film size. In particular, we find that the low
frequency excitations can be reduced in comparison with the usual
CPP configuration having a  uniform magnetization. \  Such uniform 
magnetization is susceptible to large fluctuations due to spin
momentum transfer and gives rise to appreciable $1/f$-like
noise. \cite{covington}
We investigate  DW formed by  pinning the magnetization  in the
opposite directions at the edges along the easy axis.
We show that DW motion can be controlled by a current
perpendicular to the plane of the DW magnetization in contrast
with the currently actively studied  case of  DW motion in
nano-wires. \cite{saitoh}

The LL equation  is  the basis for the present study. Since,
we are  looking at thermal and current effects, a random field and
a spin torque term are also added to the LL
equation\cite{brown,slon,rebei}.
We show that in the  spin valve (SV) geometry suggested here the spin
torque can provide the force needed to move the wall in a
controlled fashion with  $100$ times smaller current
values than those needed to switch the uniform magnetization in a 
CPP SV.

\ The paper is organized as follows. \ In section
\ref{sec:background} we describe both the 
theoretical background and the
computational model needed for our study.\ In section 
\ref{sec:modes}, we discuss the DW
structure and  we also calculate the lowest eigenmodes of the DW
using a simple one dimensional model and compare it to the
numerical solution. \ It is shown that if the center of the DW is
allowed to drift along the easy axis, the contribution to the
1/f-type noise increases. \ In section \ref{subsec:spz}, we study
the effect of an external field and a CPP current on the SV
with and without DW. \ We find  that SV with a uniform
magnetization can be susceptible to unwanted behavior due the spin
torque  driven instabilities that are absent in a 
 SV with a constrained DW.
\ We consider a new CPP geometry with a constrained DW between
  two fixed
layers one pinned along and the other pinned
 perpendicular to the direction of the
current flow. We find that  in  this CPP structure the DW motion
can be well controlled with current densities which do not lead to
the magnetization instabilities.
It is shown that accurate quantification of demagnetizing  fields
is not essential for a qualitative understanding of the influence of
the spin torque on the DW. \ In section \ref{sec:summ-conc}, we
summarize our results.

\section{Theoretical Background and Computational Model}
\label{sec:background}

For a magnetization $\mathbf{M}$ with magnitude $M_s$, the LL
equation for $ \mathbf{m} = \mathbf{M}/M_s $ with a damping in the
Gilbert form and time normalized by $\gamma M_s $, where 
$\gamma$ is the gyromagnetic ratio,  is given by
\cite{malozemoff}

\begin{equation}
\label{eq:LLG}
 \frac{d \mathbf m }{d t} = - \mathbf{m} \times
\left( \mathbf{h}^{\mathrm{eff}} + \mathbf{h}_r( t ) - \alpha \frac{d
\mathbf{m}}{d t} \right),
\end{equation}
where the effective field $\mathbf{h}^{\mathrm{eff}}$ includes the exchange
interaction, the anisotropy field along the $\mathbf{x}$ axis,  the demagnetization field, the 
Oersted field,  and
the spin torque 
\begin{equation}
\label{eq:heff}
\mathbf{h}^{\mathrm{eff}}=\frac{2A}{M_{s}^{2}}\nabla^{2}\mathbf{m}-\frac{2K}{M_{s}^{2}
}\mathbf{x}(\mathbf{m}\cdot\mathbf{x})+\mathbf{h}_{d}+p I
\mathbf{m}\times \mathbf{m}_p,
\end{equation}
The damping term is
taken to be $\alpha = 0.02$ in the absence of 
currents and is increased to $\alpha = 0.08 $ in the presence 
of spin torques to account for spin accumulation at the 
normal-ferromagnetic interface. The 
exchange constant $A=1.6
\times 10^ {-6} \; \mathrm{erg/cm}$ in this study. \ The random
field $\mathbf{h}_r(t )$ is taken to be uniform and Gaussian white
at temperature $T$, $< h_{r,i} (t) h_{r,j} (t') > = 2 \alpha k T
/( \gamma M_s^3 V) \delta_{ij} \delta(t - t')$. \cite{brown} \ In the presence
of spin torques, the white noise assumption \cite{foros}
 is strictly valid only
for frequencies around the resonant frequency as shown in Ref. 
\onlinecite{rebei}. \ Since we are only interested in currents
below the critical current, the white noise assumption will not
alter the qualitative conclusions of this work. \ The size of the
discretized cell is taken $2 \times 2 \; \mathrm{nm}^2$ in the
plane of the film. \ The inclusion of the demagnetizing field  is
important in DW motion studies and hence a numerical treatment is
often needed to get a quantitative understanding of the dynamics
of a DW \cite{schlomann,slon2,aharoni}. \ The last term in  Eq.
\ref{eq:heff} is the contribution  of  a spin torque from the pinned 
layer (PL)
 $\mathbf{m}_B$. \ The prefactor  $p$ is  dependent of
geometrical parameters  and $I$ is the current flowing
perpendicular to the magnetic multi-layers. \ The  $p$ prefactor
is dependent on  the thickness $d$, the cross section $A$ of the
layer and the polarization of the current. \ Assuming perfect
polarization of the conduction
electrons, with charge $e$,
 by the reference layer and neglecting the angular
dependence  dependence, the spin torque coefficient is given by
\begin{equation}
p  = \frac{1}{|e| d A M_s^2}.
\end{equation}
In the following simulations, the anisotropy field is taken $H_k =
200 \; Oe$ and the saturated magnetization is $M_s = 800.0 \;
emu/cc$.

\section{Excitation modes in  CPP nano-structures with domain walls}
\label{sec:modes}

\ In this section, we introduce the geometry of the DW and study
its excitation modes compared to those generated in the uniform
case.  \ We show that the in-plane components of the
magnetizations have distinctly different lowest mode frequencies
that are directly related to the inhomogeneities of the
magnetization due to the DW.
\ We also discuss the magnetization evolution  if we remove the
pinning boundary conditions and allow DW to relax to the uniform
magnetization state. In addition to the zero temperature dynamics, 
we investigate the effect of thermal fluctuations on the motion 
of transverse DW.

\subsection{Modes in the case of  constrained DW}
\label{sec:constDW}

\begin{figure}[ht!]
\begin{tabular}[c]{ll}
a &  \mbox{\epsfig{file=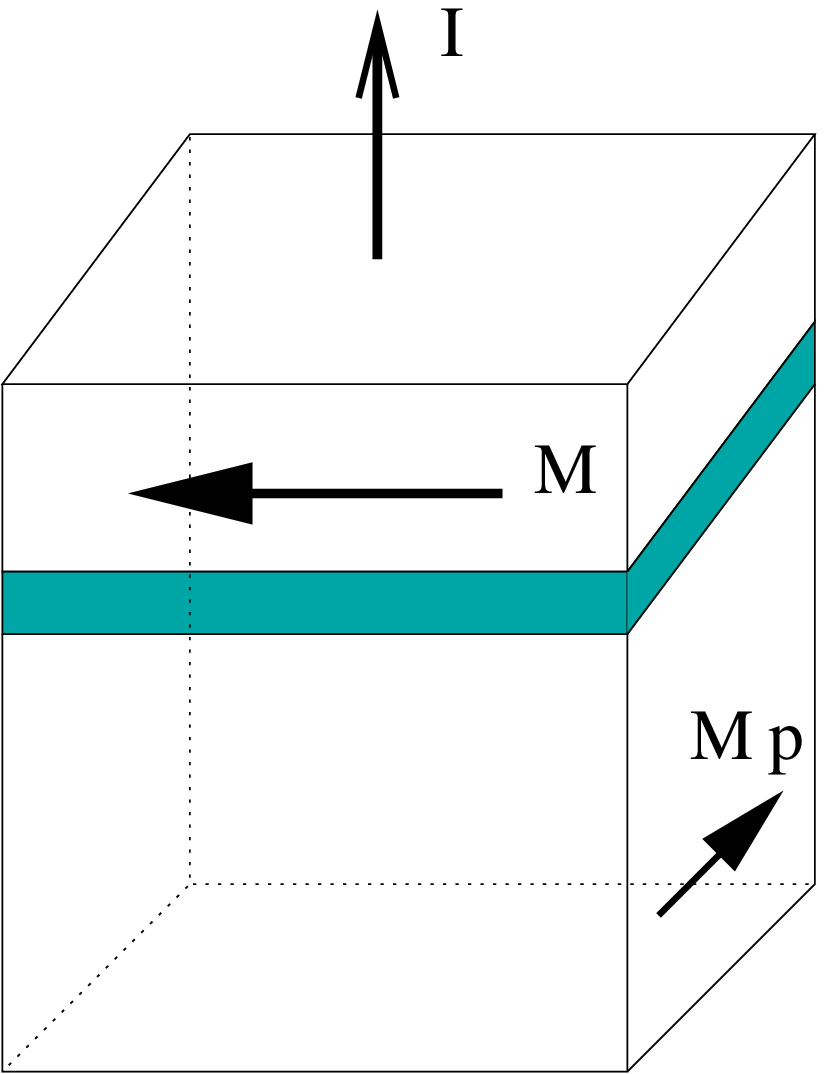,height=6 cm}} \\
b & \mbox{\epsfig{file=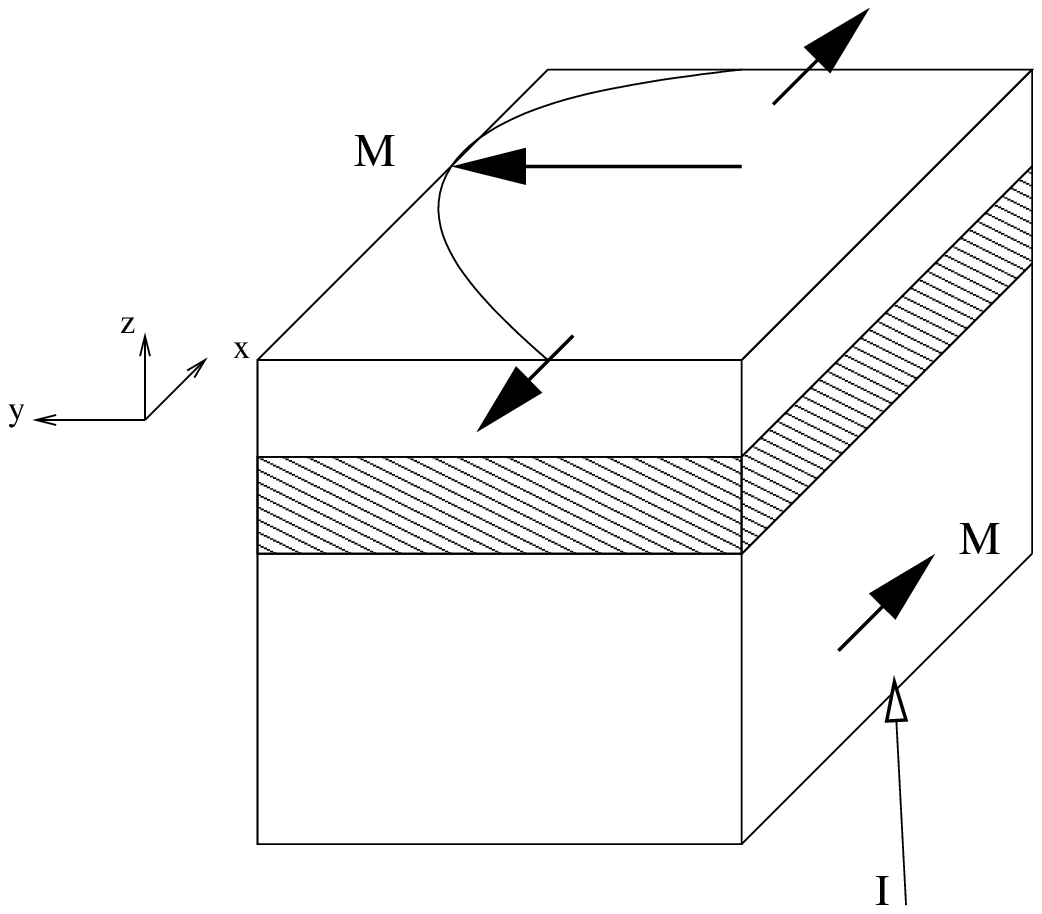,height=6 cm}}
\end{tabular}
  \caption{{{(online color) The CPP-SV magnetic nano-structures consisting
of a pinned layer (PL) with fixed magnetization  $M_p$
and free layer (FL) magnetization $M_f$ shown for cases of  ( a )
uniform magnetization and ( b) FL with tail-to tail DW. In both cases, the 
magnetization in the middle of the free layer is perpendicular to the 
magnetization in the pinned layer.
}}}
\label{fig000} \label{cpp}
\end{figure}

\begin{figure}[ht!]
 \mbox{\epsfig{file=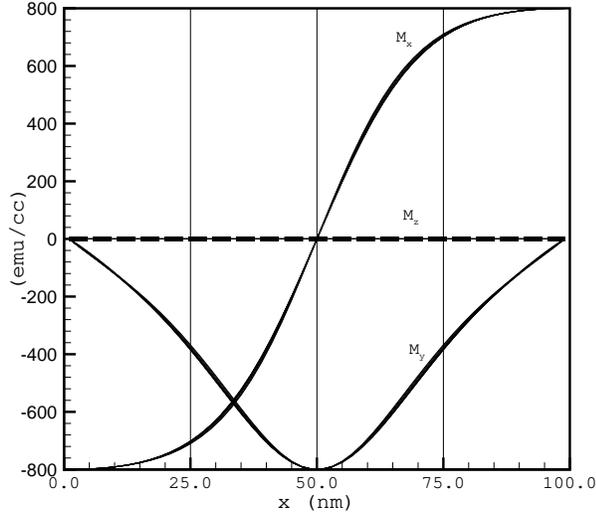,height=8 cm}}
  \caption{{{ The magnetization distribution for the three components 
of $\mathbf{M}$ in the DW ($100\times 20 \; \mathrm{nm}^2$) as a function of coordinate x along the easy x-axis.  }}}
\label{fig000c}
\end{figure}
\begin{figure}[ht!]
  \begin{center}
\begin{tabular}[c]{ll}
a & \mbox{\epsfig{file=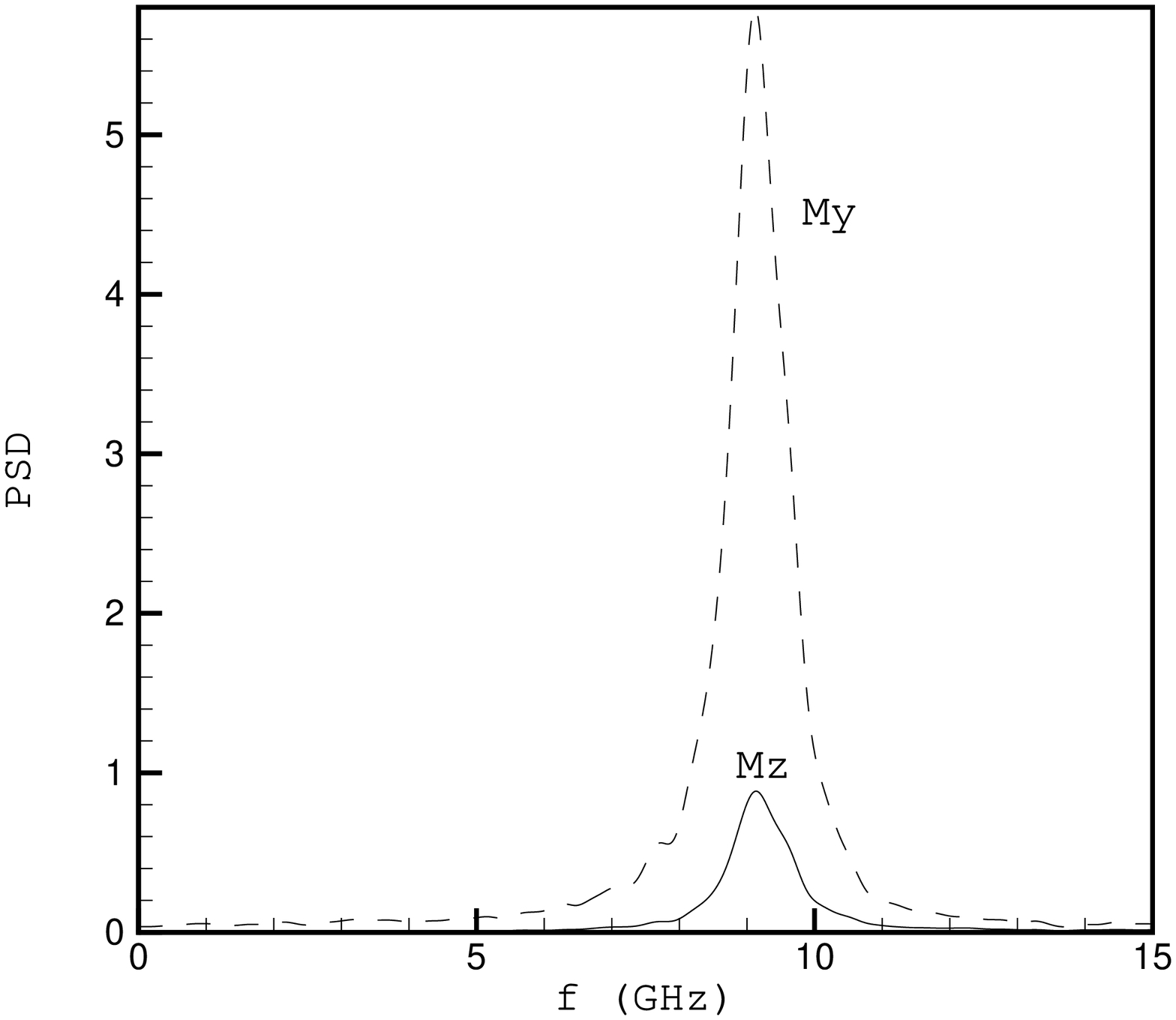,height=8 cm}}\\
b & \mbox{\epsfig{file=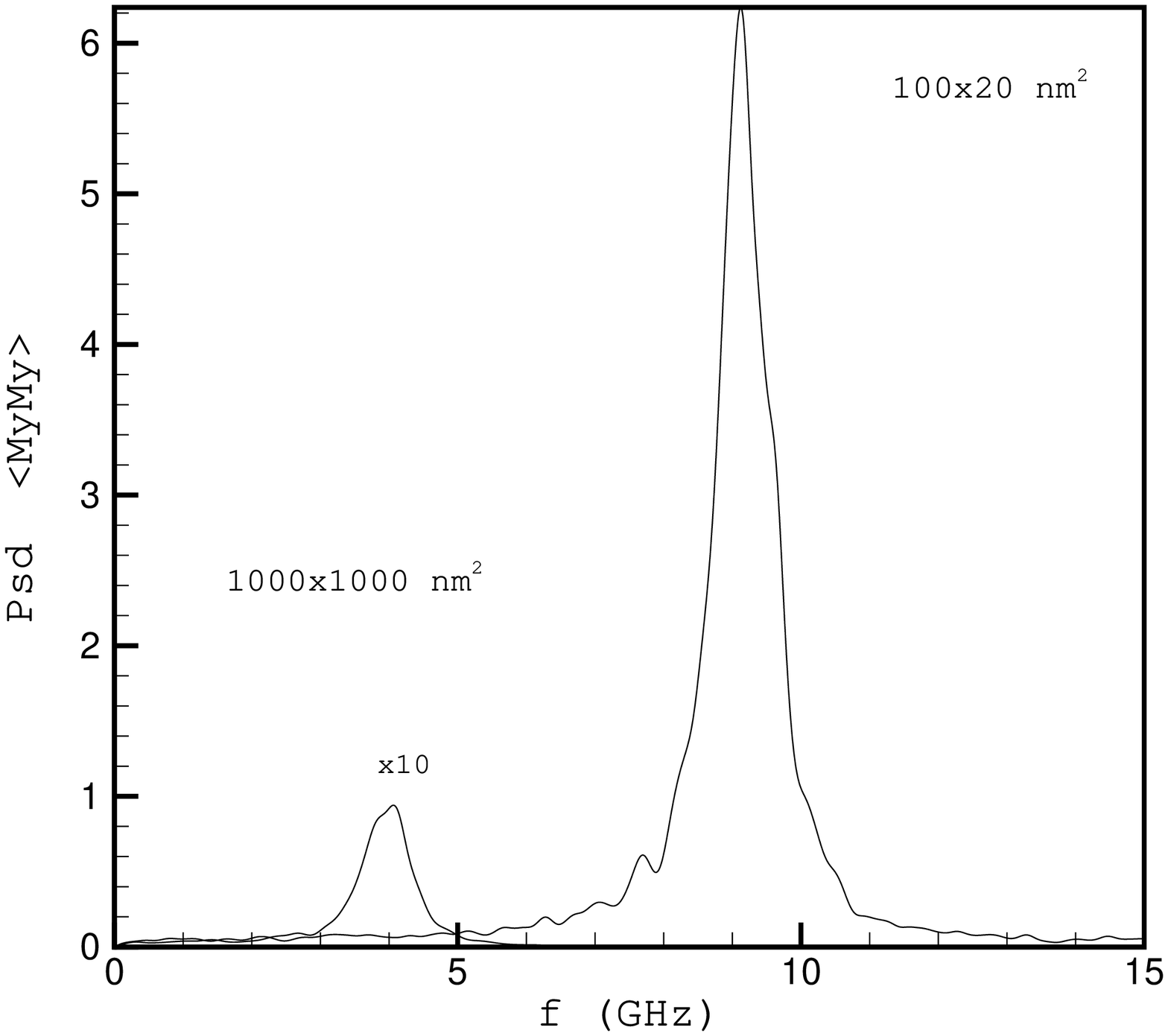,height=8 cm}}
\end{tabular}
  \end{center}
  \caption{ Spectral densities of the uniform magnetization (arbitrary units)
 in  (a)
$100\times 20$ $\mathrm{nm}^2$ and (b) $1000 \times 100$ $\mathrm{nm}^2$ thin
films as a function of the frequency (no DW).\ The
magnetization is $M_s=800 \; \mathrm{emu/cc}$ and the
anisotropy is $H_K = 200 \; \mathrm{Oe}$ along the easy x-axis. The
fluctuations of the $M_x$ component are small and do
not show up on the  scale adopted here. }
\label{fig11}
\end{figure}


\ Figure \ref{fig000} shows the geometry of the systems we have
studied. \ As can be seen from the schematic  illustration in
Fig. \ref{cpp} that in contrast with the traditional CPP-SV, we
investigate free layer (FL) with inhomogeneous magnetization 
due to DW in FL
coupled to PL  with uniform magnetization.
 \ We will add later in
section III a third magnetic layer when we discuss the effect of
spin torques on the DW.\ The film has an in-plane easy x-axis
along the direction of the magnetization of the bottom pinned
layer. \ The magnetization of the PL is taken homogeneous. \
Fig. \ref{fig000c} shows the  magnetization profile  for the case of
a small current density   with DW  formed in the plane of FL
due to the uniform pinning at the x=$\pm$ L/2 boundaries, where
$L$ is the length of the side along the easy axis.


 \ Figure \ref{fig11} shows the power spectral density (PSD) $ |
M_{i}( \omega )|^2 $
 of the FL magnetization  which has a peak at
around $9.0 \; \mathrm{GHz}$. \ Before calculating the PSD, we first
average $\mathbf{M}(x,y)$ over space.  \ The  size of the FL film
is taken to be  $100\times 20\times 2 \; \mathrm{nm}^3$ while that of the
FL is $100\times 20\times 9 \; \mathrm{nm}^3$.
\ For the unpinned boundary case, the average magnetization points
along the easy x-axis and the transverse components are
oscillating   with a frequency approximately twice the
ferro-magnetic resonance
 (FMR) frequency of an infinite
thin film given by the Kittel formula,
 $ \omega = \gamma \left( H_k \left(
H_k + 4 \pi M_s \right) \right)^{1/2} $, i.e., around $4.1 \;
\mathrm{GHz}$. \ Figure \ref{fig11} shows that a film-size of $1000 \times
1000\times 3 \; \mathrm{nm}^2$ has practically the same FMR peak as that of
an infinite thin plate. \ Thus,  the presence of boundaries is an
important factor which will be discussed throughout the rest of
the paper.

\ First we start discussing  the DW relaxation as we remove
boundary pinning. This relaxation process reflects intrinsic DW
modes.
\ Figure \ref{fig001} shows the
relaxation of the DW when the pinning at the edges is removed at
$t = 0 \; \mathrm{ns}$. \ The average x and z components stay zero for more
than $0.1 \; \mathrm{ns}$ after turning off the pinning. \ Afterward, the
x-component converges to $M_s$ and the z-component begins
oscillating around zero. \ The y-component starts oscillating
immediately around a non-zero average after the removal of the
pinning at the edges and after $0.1 \; \mathrm{ns}$ starts oscillating
instead around zero. \ The initial phase of this decay of the DW
to the uniform state shows interesting features. \ The x-component
shows
  a {\it compression-decompression} mode
which  represents
 oscillations of the DW around the center and along
the easy x-axis (see Fig. \ref{fig001}b-c). \ Simultaneously, the
y-component shows a behavior similar to
 a {\it breathing} mode. \ Finally the last plot shows how a uniform
magnetization which is initially along the hard axis relaxes to the
state along the easy axis. \ Figure \ref{damping} shows that large
damping makes the DW more stable to external perturbations.

\begin{figure}[ht!]
  \caption{Transient dynamics from the DW state to the uniform magnetization state
  with thermal fluctuations included:
(a) magnetization components a a function of time when pinning has
been turned off  at $ t = 0 \; \mathrm{ns}$. (b) shows the profile of the
$M_x$ component at different time steps $1, 16, 33, 50$ and $60$ in units of
$\Delta t = 0.05$ ns. \ (c) same
as in
 (b)but for the $M_y$ component. (d)
Relaxation of the uniform magnetization from the hard axis
position to the easy axis position compared to that of the case of
the DW. The damping constant is $\alpha = 0.02$.\ The width of the
curves in (b) and (c) represent variations in the y direction at a
given point along the x-axis.(this figure can be obtained from one of the authors directly:arebei@mailaps.org or check the published version in Phys. Rev. B June issue 2006) } 
\label{fig001}
\end{figure}


\   Magnetization dynamics  of the DW can be characterized in
terms of its normal modes. In \ Fig. \ref{fig10} we show the
spectral densities of the different components of the
magnetization found in  the DW configuration. \ In this case, the
x component has a peak at lower frequency than the
y-component. \ This higher frequency is
directly due to the boundary conditions on the magnetization. \
The dependency of the magnetization on the y coordinate appears to
be very weak. \ As a function of the position x, the x and z
components have a configuration which is odd under reflection with
respect to the
 center, while the y component
configuration is even.

\ These different excitations of the DW can be qualitatively
understood in a 1-D calculation with a  simple approximation for
the demagnetization field that of an infinite thin film. \ If we
take, $\mathbf{m}=\left(
\sin\theta\cos\varphi,\sin\theta\sin\varphi,\cos \theta\right)  $,
then the equations of motion for the angular variables are given by
\begin{equation}
\frac{d\theta}{dt}=h_{y}^{\mathrm{eff}}\cos\varphi-h_{x}^{\mathrm{eff}}
\sin\varphi-\alpha
\sin\theta\frac{d\varphi}{dt}, \label{xeq}
\end{equation}
\begin{equation}
\sin\theta\frac{d\varphi}{dt}=h_{z}^{\mathrm{eff}}\sin\theta-\cos\theta\left(
h_{x}^{\mathrm{eff}}\cos\varphi+h_{y}^{\mathrm{eff}}\sin\varphi\right)  +\alpha\frac{d\theta
}{dt}.\label{yeq}
\end{equation}
\ We are looking for excitations around the ground state. \ If we
take  the magnetization to be in-plane, i.e, $ \theta_{0}
=\frac{\pi}{2}$, and depends only on the x-coordinate, we find that the
static solution should  satisfy  the Sine-Gordon equation
\begin{eqnarray}
\frac{d^{2}\varphi_{0}(x)}{dx^{2}}+\frac{1}{\lambda^{2}}\sin\left[
2\varphi_{0}(x)\right]   =0,
\end{eqnarray}
with the boundary condition
 $
\varphi_{0}\left( - L/2\right) = \pi
$ at the left edge and $\varphi_{0}\left( L/2 \right) = 0$  at the right edge.
 \ $ \lambda=\sqrt{2A/K}  $ is the width of
the DW. \ It should be noted here that the same condition arise for DW 
in infinite films and the
only difference 
 is in the boundary conditions. \ An analytical solution
for this equation does not appear to be possible but it can be found
numerically. \cite{houston} \ After linearization,
$\theta \rightarrow
\theta_0 + \theta, \; \;  \varphi \rightarrow \varphi_0 + \varphi$,
the equations of motion, Eqs. \ref{xeq} and \ref{yeq}, become
\begin{eqnarray}
\frac{d\theta}{dt}  & = & \frac{2A}{M_{s}^{2}}\frac{d^{2}\varphi(x)}{dx^{2}%
}+\frac{2K}{M_{s}^{2}}\varphi\cos2\varphi_{0}-\alpha\frac{d\varphi}{dt}, \\
\frac{d\varphi}{dt}  & =& -\frac{2A}{M_{s}^{2}}\frac{d^{2}\theta(x)}{dx^{2}%
}+G\left(  x\right)  \theta-\alpha\frac{d\theta}{dt},
\end{eqnarray}
where the function $G$ is
\begin{equation}
G(x)=-\frac{2A}{M_{s}^{2}}\left(  \frac{\pi}{L}\right)  ^{2}+4\pi-\frac
{2K}{M_{s}^{2}}\cos^{2}\left(  \frac{\pi x}{L}\right).
\end{equation}
\ To obtain the function $G$, we have Fourier transformed $\varphi_0(x)$ and kept only the first term.
\ This is sufficient to understand qualitatively the main results
of the simulation.

\ The time-dependent variables $\theta$ and $\varphi$
 satisfy homogeneous boundary conditions and represent fluctuations
around the equilibrium solution. \  If $\varphi(x,t)=\varphi(x) \exp(i \omega t)$,
then $\varphi(x)$ can be
 written in the following form to satisfy the boundary conditions

\begin{equation}
\varphi(x)=\sum_{n=1,3,5,...}a_{n}\cos\left(  \frac{n\pi x}{L}\right)
+\sum_{m=2,4,6,...}b_{m}\sin\left(  \frac{m\pi x}{L}\right),
\end{equation}
where $x$ varies in the range $ - L/2 \le x \le L/2 $.
\ The equations of motion then become algebraic equations in $a_n$ and
$b_n$. \ Then within a  linear approximation, the magnetization components
are given by
\begin{eqnarray}
m_{x}  & = & \cos\varphi_{0}(x)-\varphi(x)\sin\varphi_{0}(x), \\
m_{y}  & =& \sin\varphi_{0}(x)+\varphi\cos\varphi_{0}(x).
\end{eqnarray}
\ Since the normal frequencies of the system depend on the
wavenumber $n \pi/L$, we see that because of the parity
of the ground state $(m_x^0=\cos \varphi_0,m_y^0 = \sin \varphi_0)$, the
lowest wavenumber
that appears in the y-component is larger $(m=2)$ than that of the
x-component $(n=1)$. \ This is the reason why the breathing mode has
a higher frequency than the spring (or Doring-like mode)
mode, Fig. \ref{fig10}-\ref{fig00x}. \ In an infinite plane, the spring mode becomes
the Doring mode in our case. \ The Doring mode is associated with translation
of the DW, i.e., with a zero frequency mode. \ In a constrained DW, the pinned edges provide a restoring force
and hence the center of the DW will oscillate instead of translating.

\begin{figure}
\mbox{\epsfig{file=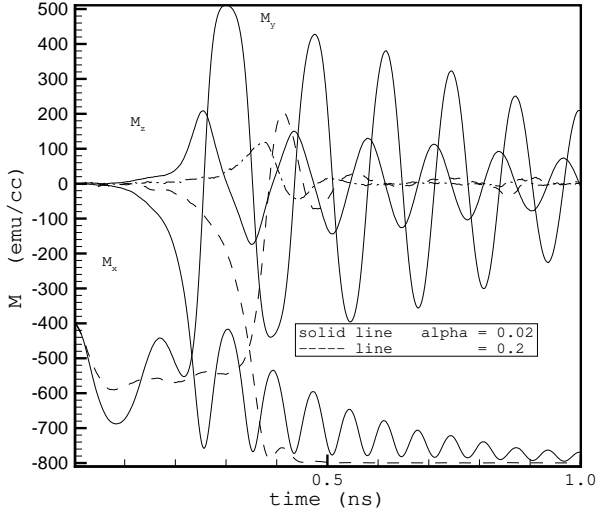,height=8 cm}}
\caption{Relaxation of the DW  to the uniform state as a function of damping for $\alpha = 0.02$ and
$\alpha = 0.2$. }
\label{damping}
\end{figure}

\begin{figure}[ht!]
  \begin{center}
\mbox{\epsfig{file=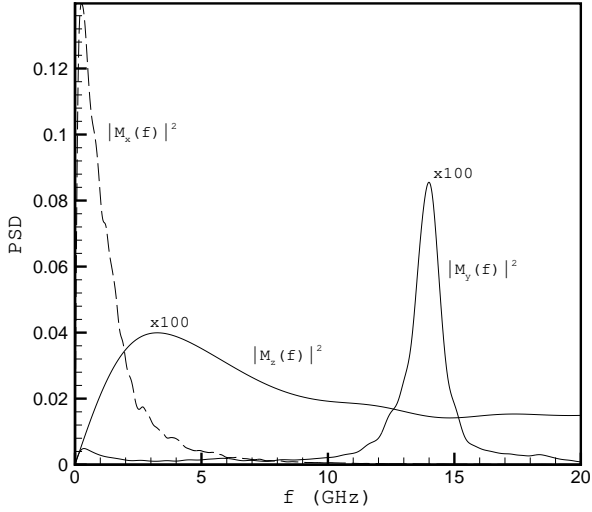 ,height=8 cm}}
  \end{center}
  \caption{ Spectral densities of the magnetization in the constrained DW (arbitrary units).
  The x  component has a peak around, $f = 1.0 \mathrm{GHz}$.
  This frequency is that of the Doring-like mode in a finite geometry.
  The peak of the y-component is at $f=14 \; \mathrm{GHz}$ which
corresponds to the breathing mode.
   Both amplitudes for y and z have been
multiplied by 100 for better visualization.}
\label{fig10}
\end{figure}

\begin{figure}[ht!]
  \begin{center}
\begin{tabular}[c]{ll}
a & \mbox{\epsfig{file=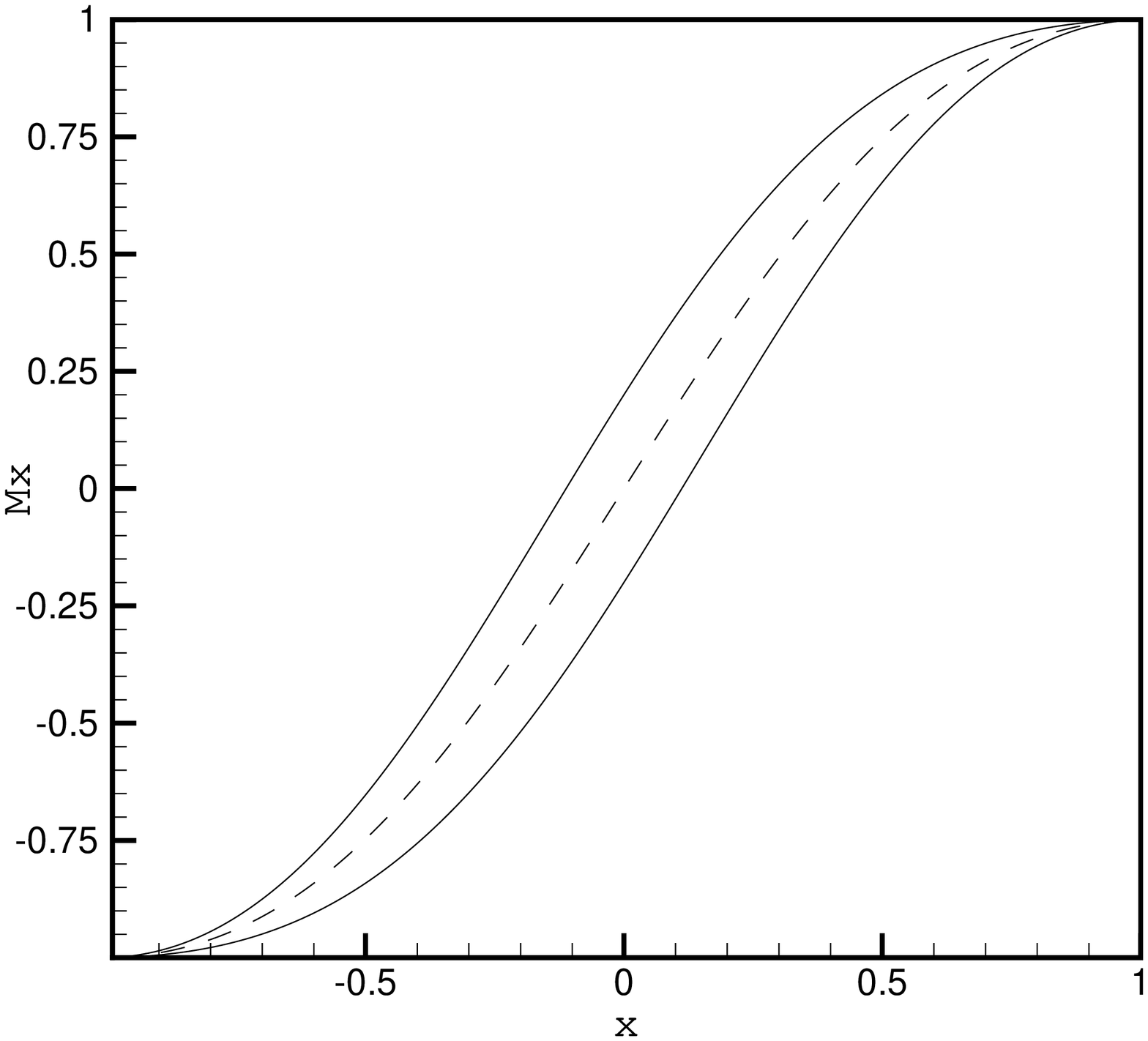,height=8 cm}} \\
b  & \mbox{\epsfig{file=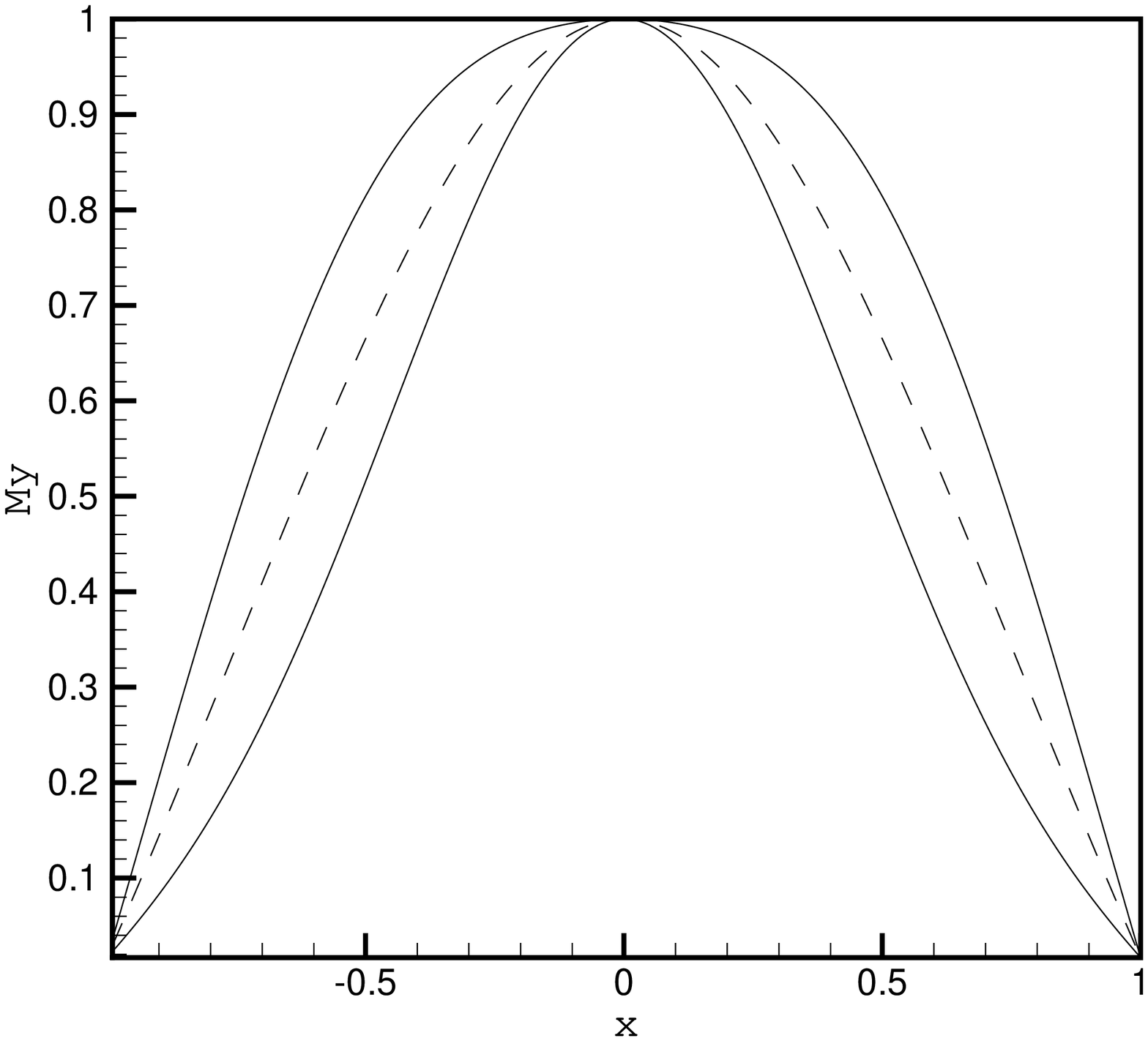,height=8 cm}}
\end{tabular}
  \end{center}
  \caption{Lowest modes of the DW: The 'spring' mode (a) and the 'breathing'
mode (b) of the DW. The dashed line is the equilibrium solution.
The solid curves represent the amplitudes of the modes around the equilibrium state. \ Both $x$ and $M_{x,y}$ have been normalized.}
\label{fig00x}
\end{figure}

\ The breathing mode is different from both the Doring and
  Winter modes. \ The Winter modes exist only in
 in infinite DW as opposed to the
 constrained DW treated here. \ As it can be seen
from Fig. \ref{fig001}c and Fig. \ref{fig00x}b, the 
breathing mode is the mode
that is mostly excited when the pinned boundary conditions are turned off.

\subsection{Low frequency noise due to a drifting domain wall}

\ Finally in this section, we discuss the advantages behind
constraining the DW to regions comparable 
in size to the DW width. \  We show that DW motion in 
large films can be the origin
of $1/f$-type noise. This noise has
already been  suggested from the  measurements in Refs.
\onlinecite{ingvarsson,philipp} and is detrimental to any 
sensing device.

\ Indeed we find that $1/f$-type noise increases if  we increase the
size of the thin film so it is much larger than the wall width
$\lambda$. \ Hence in this case, the center of the DW is allowed
to drift away from the middle of the film in either direction
along the easy axis due to thermal fluctuations and 
the demagnetization field. \ It has been
suggested in many experiments that DW motion can give rise to
$1/f$-type noise \cite{hardner}. \ Low frequency noise usually makes
structures with non-uniform magnetization undesirable for use in
magnetic sensors. \ However, the magnetization dynamics pattern
changes dramatically if we constrain the DW. \ In the following,
 we first show that making the DW unconstrained  does indeed lead
to the  $1/f$-type noise in general agreement with experimental
results \cite{ingvarsson,philipp}.

\begin{figure}[ht!]
  \begin{center}
  \end{center}
  \caption{{{Magnetization Dynamics in the $320
\times 20 \; \mathrm{nm}^2$ elongated CPP valve in the case of $ L= 320 \; \mathrm{nm} \gg \lambda$; $\lambda$ is the DW width :
 \ (a) The average x-component
of the magnetization shows dynamics that gives rise to telegraph
noise. The DW moves in a thin long strip due to random thermal
excitations. \ (b) Spectral densities in the x-, -y and
z-components of the magnetization in the elongated strip (arbitrary units).
Substantial low frequency telegraph type noise
is apparent in the x-component. The amplitudes of the y and z
peaks have been magnified 100 times.
\ (c-d)  Magnetization distribution in the $xy$-plane at two different 
times (c) and (d) that correspond to states with positive and negative average 
$M_x$, respectively. This figure can be obtained from the first author or check out the published version in Phys. Rev. B. June 2006.
 }}} \label{fig5} \label{fig3}
\label{fig4}
\end{figure}
\begin{figure}[ht!]
\begin{tabular}[c]{llll}
 \mbox{\epsfig{file=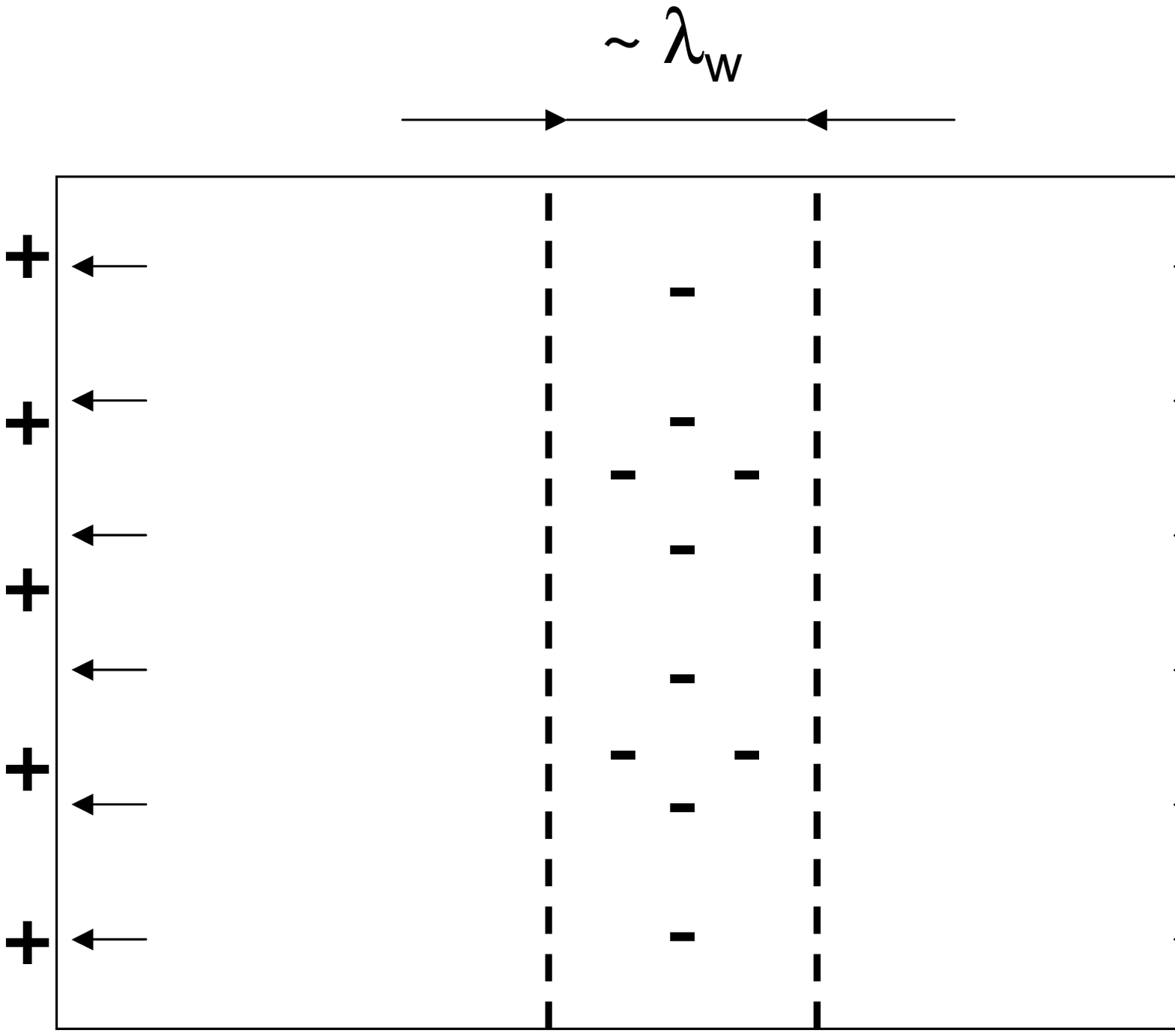 ,height=7cm}}
& \mbox{\epsfig{file=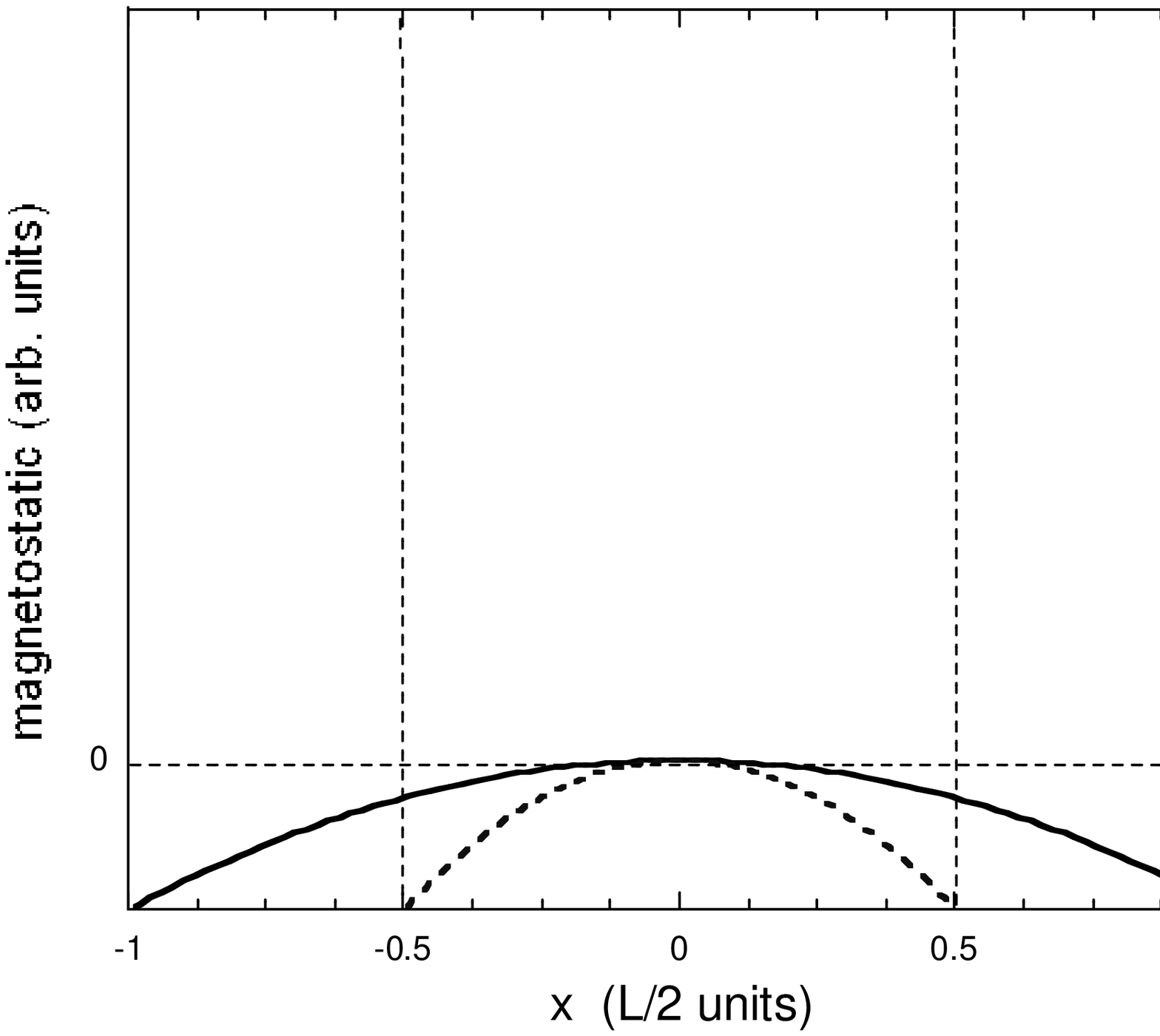 ,height=7cm}}\\
\mbox{\epsfig{file=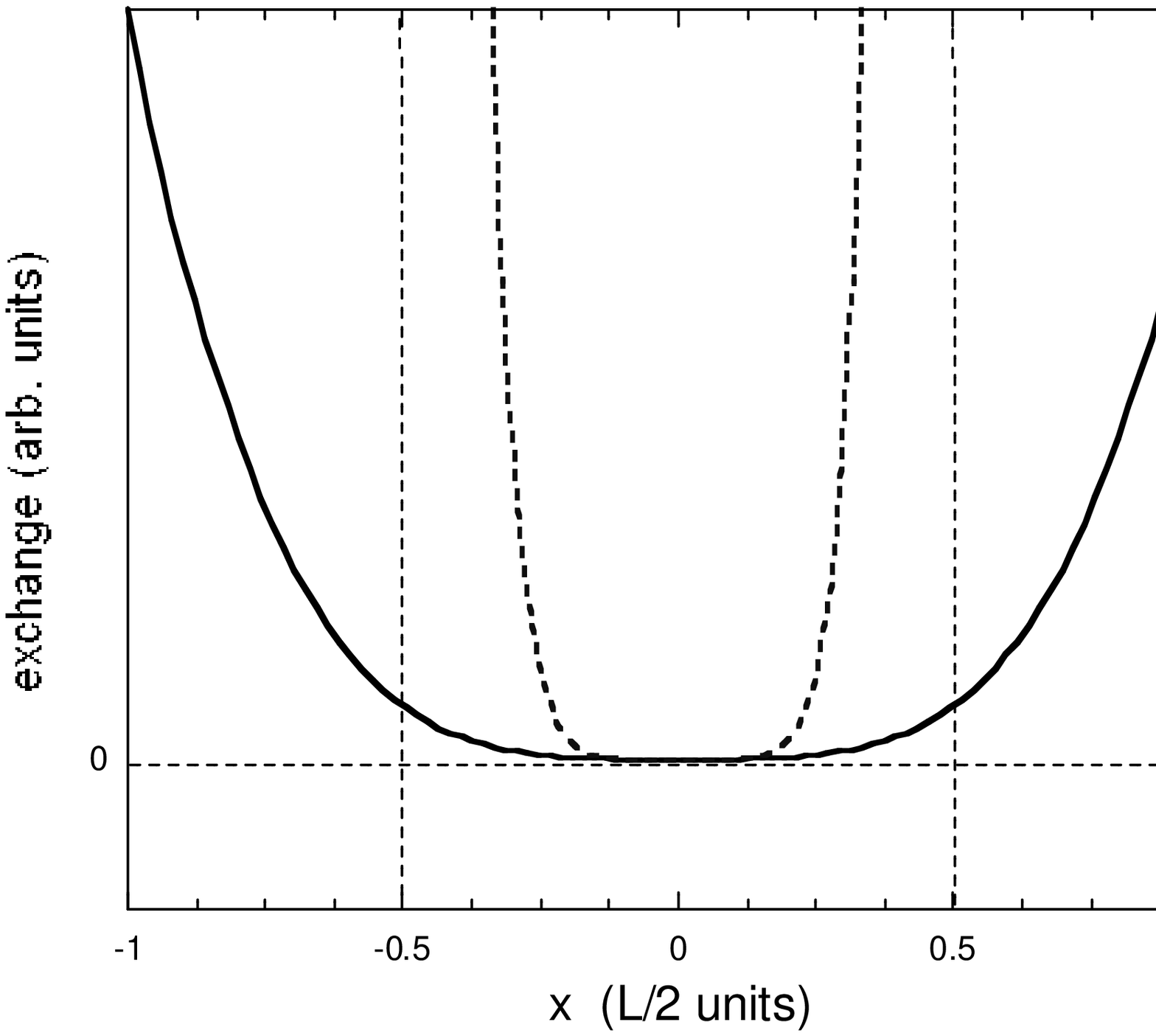 ,height=7cm}}
& \mbox{\epsfig{file=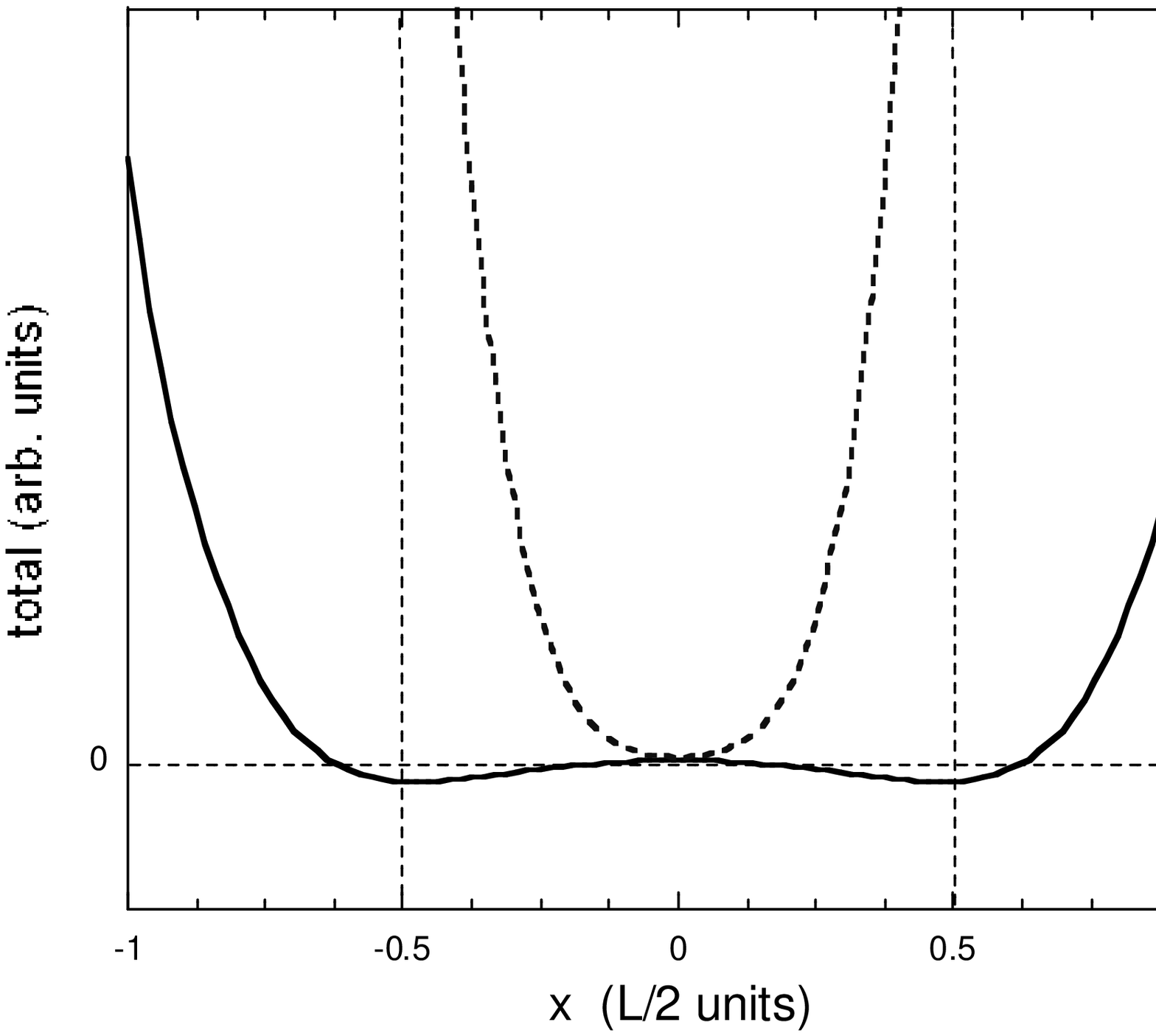 ,height=7cm}}
\end{tabular}
\caption{
 Schematic illustration of a)  magnetic charge distribution and
 contributions to  DW potential due to  b) corresponding magnetostatic
energy
 c) exchange energy and  d) total potential for  DW motion along
 the easy  axis. The DW, with width $\lambda_{w}$, is shown to acquire negative charges
due to $\nabla \cdot \mathbf{M}$ distributed around DW center and
 positive charges at the edges. The dashed line is used for the
 case of  constrained DW ($\lambda_{w}$ $ \approx $ L) and contrasted with  the case of
 unconstrained DW (solid lines) with $\lambda_{w}$ $ << $ L.
  }
\label{demag}
\end{figure}


\ To observe low frequency behavior in this system we need to
reduce the effect of the restoring force on the DW. \ This amounts
to enlarging the length of the sides along the easy axis of the
film to be much larger than the width of the DW. \ This way the
{\it whole} DW can move from the left to the right and back
because of thermal fluctuations ( Fig. \ref{fig5}). \ Such
behavior has been observed in many experiments
\cite{hardner,ingvarsson}. \ Figure \ref{fig3} shows the PSD's
associated with the different components of the magnetization. \
The x-component shows the $1/f$-type behavior as expected.  \ Figure
 \ref{fig4} shows  a real-time trace for the average components of
the magnetization. \ The x-component shows 'switching'-type
behavior between two states, a signature of telegraph noise. \ The
remaining two components are very stable and have much higher
frequencies. \ The $M_x ( t )$ behavior of the magnetization
resembles the evolution of a two state system (telegraph noise). \
It is clear from Fig. \ref{fig4} that this telegraph noise
originates from the DW motion in a shallow double well potential.
 Finally,  it is apparent from  Fig. \ref{fig4} that DW
magnetization appears to spend more  time  closer to the boundary
than in the center. This behavior coupled  
with  the   telegraph-like noise
  are indicative of the  double well potential for motion in
the case of 'unconstrained' DW.\ 
In the case of constricted DW,  we do not find this double
well potential  instead the potential has a single minimum 
and this is 
the reason behind the different noise features observed 
in both cases. \ However, it is not obvious what is the origin of
such a double well potential and  why it  disappears in the case
of a constricted DW.

 \ In the following we offer a simple explanation for such 
a behavior. \cite{berger}  \
This difference between unconstrained and constrained DW's can be
understood using the notion of charged DW introduced by Neel
\cite{Neel_dw}. \ The DW in elongated nano-elements belongs to this
category and is characterized by magnetic charges distributed in the
vicinity of the DW center as schematically  shown in Fig.
\ref{demag}(a). \ Moreover because of the finite size of the plate,
there are
significant boundary charges at either end of opposite sign to that in the 
DW. \ Thus
magnetostatic interaction associated with these charges leads to
a potential as a function of DW position as shown in
Fig.~\ref{demag}(b). The exchange interaction contribution shown
in Fig.~\ref{demag}(c) has very strong  size dependence due to
large exchange energy increase as DW  approaches pinned
boundaries.\ Thus, it becomes clear that the total potential for  DW
motion for unconstrained DW has a double well feature which
disappears as the DW gets constrained as shown in Fig. \ref{demag}(d). In 
our case the height of the barrier between the two wells is less 
than $k T$.

\clearpage

\clearpage

\section{Effect of spin torques and Zeeman terms on the DW}

\label{subsec:spz}

\ In this section we investigate the effect of spin currents and
external fields on the magnetization in SV.

\ First we  study  a uniformly
magnetized SV and calculate
the spectral density of the x-component of
the magnetization. \ We show that the
spin torque can be a source of instabilities in this case. \ However in
the DW case, we show that the effect of the spin torque
can be used instead to control its motion. \ The CPP structure where
this is possible
 is different than previously proposed structures. \ We instead add another
magnetic layer to polarize the current in the direction {\it perpendicular}
to the plane of the SV.

\subsection{Noise in a CPP spin valve with uniformly
 magnetized layers}

\begin{figure}[ht!]
  \begin{center}
 \mbox{\epsfig{file=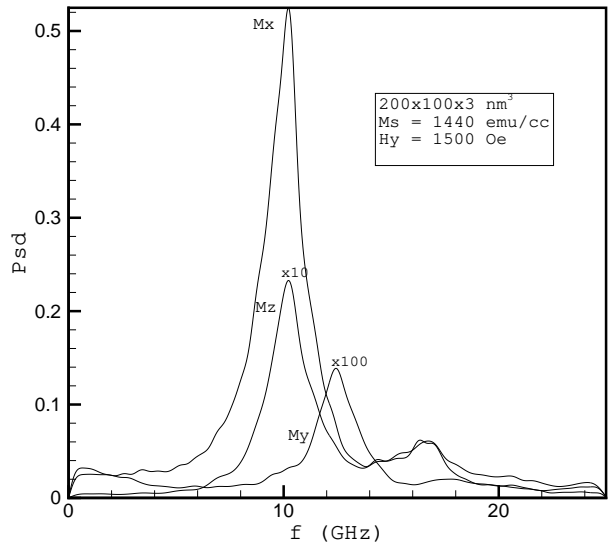,height=8.5 cm}}
  \end{center}
  \caption{Spectral densities (arbitrary units) of the magnetization in the free layer in the absence of current.}
\label{200x100}
\end{figure}

\label{subsec:noisecpp}

\ Our discussion here will be closely
related to the experimental findings in reference
 \onlinecite{covington} where it was shown that spin  transfer
in a CPP device can give rise to 1/f-type noise.\ The noise range
can be in the GHz regime and in effect makes the use of a CPP device
as a GMR sensor unattractive.

\ In the following we discuss a
SV similar  to the one 
treated in Ref. \onlinecite{covington} where the magnetization of the
free layer is perpendicular to the pinned magnetization. We use
 a single spin picture to discuss the noise 
in this system. \ We
show that this model can reproduce to a great extent the 
trend in the noise
spectrum observed in the experiment in Ref.
 \onlinecite{covington}. \ Adopting a single
particle picture could be a rather crude approximation in this
case \cite{rebei,rebei2}, but it is sufficient  for our purpose to
demonstrate the contribution of the spin torque to the  noise of a
CPP device observed in Ref. \onlinecite{covington}. \ Moreover, the
single domain picture discussed here will help us in the
interpretation of the numerical results of the more involved case
of a DW.

 \ The CPP-SV with uniform magnetization
is shown in Fig. \ref{cpp}a. \ We take the effective
field  to be equal to $\mathbf{H}^{\mathrm{eff}}= \left( H_b -
\nu H_{c}, 1500, -4\pi M_{z}\right)  $\ $\mathrm{Oe}$ where
the $x$-component is much
smaller than the $y$ component. \ Therefore in this case the
magnetization is expected to be almost perpendicular to the one of
the pinned layer. \  The saturated magnetization
is equal to $1500$
$\mathrm{emu/cc}$. \ The AFM field from the pinned layer
is assumed small $H_{c}=20 \; \mathrm{Oe}$\ and the constant
$\nu =1$\ if the pinned magnetization is along $+x$\
and $\nu=-1$\ \ if the pinned
magnetization is pointed in the $-x$\ direction. \ The spin
torque term will be
represented by an 'effective' field term $\mathfrak{p}=$\ $1000 \; \mathrm{Oe}$\
which is equivalent to having a $10 \; \mathrm{mA}$ fully polarized
current flowing
into the free layer. \ These
parameters are chosen to be close to those
used in the experiment of reference \onlinecite{covington}. \ Using
similar parameters, figure \ref{200x100} shows the PSD for the 
magnetization
in a $200 \times 100 \times 3 \; \mathrm{nm}^3$ thin film. \ This 
micromagnetic calculation  
clearly shows that magnetization is
almost uniform and is  closely
aligned with the $1500 \; \mathrm{Oe}$ field along the y-axis. Hence 
we can use a macro-spin picture to calculate the noise 
spectra in this system.

First, we need to determine the equilibrium position in 
the presence of the
spin torque which is not always possible. The spin torque 
here  is comparable to the
precession torque from the effective
field. \ The equilibrium state is found  by solving the simultaneous
equations
\begin{eqnarray}
M_{x}  & = & r\left(  H_{e} - \nu H_{c}\right)  \\
M_{y}  & = & r\left(  H_{y}- \nu\frac{\mathfrak{p}}{M}M_{z}\right)  \\
M_{z}  & = & r\left(  -4\pi M_{z}+ \nu\frac{\mathfrak{p}}{M}M_{y}\right)
\end{eqnarray}
with the constraint $M_{x}^{2}+M_{y}^{2}+M_{z}^{2}=M_s^{2}$ and
$r$\ is a real number to be determined. \ These algebraic
equations usually have up to four solutions and hence a stability
analysis is needed to determine the stable solutions. \ This will
be part of the PSD calculation of the x-component of the
magnetization. \ Once, we have found the static solution(s)
$\mathbf{M}_{0}$, we make a linear expansion around it,
$\mathbf{M}=\mathbf{M}_{0} + \mathbf{m} (t)$, where the
perturbation is assumed to have the form
$\mathbf{m}(t)=\mathbf{m\exp}\left(  -i\omega t\right)  $. \ The
noise is calculated by calculating the susceptibility or the
linear response of the magnetization due to an external small ac
field $\mathbf{h} (t)$. \ This argument neglects the fact that
establishing a current across the layers is a non-equilibrium
process and that a fluctuation-dissipation argument such as the
one used below is not valid in general. \ However we have
shown in Ref. \onlinecite{rebei} that for a system in
quasi-equilibrium, deviations from the equilibrium fluctuation
dissipation relation are significant only for frequencies far from
the FMR frequency of the system. \ We assume in the following that
the noise in our model depends only on the equilibrium state of
the magnetization and hence only the noise around the FMR peak is
well described by the method adopted here.

\begin{figure}[ht!]
  \begin{center}
 \mbox{\epsfig{file=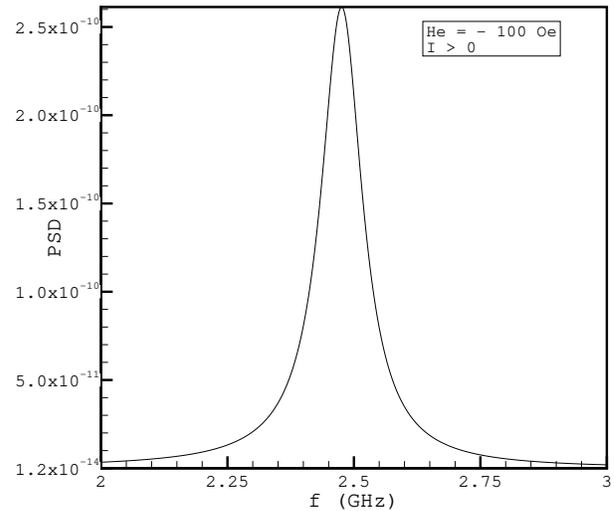,height=8 cm}}
  \end{center}
  \caption{{{The FMR curve for the free layer with uniform magnetization and for negative bias fields (arbitrary units). }}}
\label{100n}
\end{figure}

To solve for the small perturbations from equilibrium, we need to solve the
following system of equations,
\begin{equation}
\left(  i\omega \mathbf{I}+ \mathbf{A}\right)  \cdot\mathbf{m}=\mathbf{d,}\label{m}%
\end{equation}
where the coefficients of the matrix $\mathbf{A}$ are determined from the equations
of motion for the magnetization,

\begin{eqnarray}
\mathbf{A}_{11}  & =&\frac{\alpha\gamma}{M_s}\left(  -H_{y}M_{0,y}+4\pi M_{0,z}^{2}\right)
,\nonumber\\
\mathbf{A}_{12}  & = &\frac{\alpha\gamma}{M_s}\left(  2H_{e}M_{0,y}-H_{y}M_{0,x}\right) \nonumber \\
&& + {\frac{\gamma}{M_s}}\left(  -2 \nu \mathfrak{p}M_{0,y}+4\pi M_s M_{0,z}\right)  ,\nonumber\\
 \mathbf{A}_{13} & = &\frac{2\alpha\gamma}{M_s}\left(  H_{e} M_{0,z}+4\pi M_{0,x}%
M_{0,z}\right) \nonumber \\ 
&& +{\frac{\gamma}{M_s}}\left(  -2 \nu \mathfrak{p}M_{0,z}+4\pi
M_s M_{0,y}+H_{y}M_s\right)  ,\\
d_{1}  & = & -\frac{\alpha\gamma}{M_s}\left(  M_{0,y}^{2}+M_{0,z}^{2}\right)
,\nonumber
\end{eqnarray}
\ The remaining coefficients can be determined in a similar way.

\ The coefficients in front of the x-component of the
ac field $\mathbf{h}$(t)
are grouped in the vector $\mathbf{d}$. \ The stable
solutions will be those for which the imaginary part of $\omega$ is
negative or zero, $\det\left|  i\omega + \mathbf{A}\right|  =0.$ \ In the
absence of
the spin torque, the frequencies are  real in a stable
system. \ The imaginary
frequencies that appear are a signature that the spin torque
can act as a (damping) force.\ The noise
spectrum  is
found by solving for $\mathbf{m}$ in Eq. \ref{m}. In the experiment only the
noise in x-component, $C_{xx}(\omega)=\int d t \langle M_x(t)M_x(0)\rangle e^{i \omega t} $
along the pinned magnetization is of interest.\ It is
found
from  the fluctuation-dissipation relation at inverse temperature $\beta$
\begin{equation}
C_{xx}\left(  \omega\right)  =\frac{1}{\omega}\mathrm{Coth}\left(  \frac{\beta\omega
}{2}\right)    \operatorname{Im}   \frac{\det\left|
\begin{array}
[c]{ccc}
d_{1} & A_{12} & A_{13}\\
d_{2} & i\omega+A_{22} & A_{23}\\
d_{3} & A_{32} & i\omega+A_{33}%
\end{array}
\right|  }{\det\left|  i\omega+ \mathbf{A} \right|  }.
\end{equation}
These steps are carried out for all
the static solutions that are   found for each bias field $H_e$
in the presence of the
spin torque. \

\ The magnetization of the pinned layer is taken in the -x direction (as in
the experiment) and
the current is positive when it flows from the pinned to the free
layer. \ In this case we expect to see more noise for negative easy axis
fields and less noise for positive easy axis fields. \ The $1/f$-type noise
is
observed  when the field along the easy axis
is small and negative. \ Since this equilibrium
analysis can not show actual
switching between two
states as in the simulations \cite{rebei2} and
the experiment, we may be able to
deduce the switching indirectly since depending on the value
of the $H_e$ field, we may end up with  more than one possible
solution to the static equations.
\ For large negative easy axis fields (Fig. \ref{100n}), we see the usual
 shape of  FMR
curves. \ The PSD curves in this section only are normalized differently
from those in other sections of the paper. \ The  damping parameter
in this calculation is taken $\alpha = 0.005$, which is appropriate
for a permalloy even though we expect a higher value due to spin accumulation at
the interfaces between a normal conductor and a ferromagnet. 
 \cite{rebei}

\begin{figure}[ht!]
  \begin{center}
 \mbox{\epsfig{file=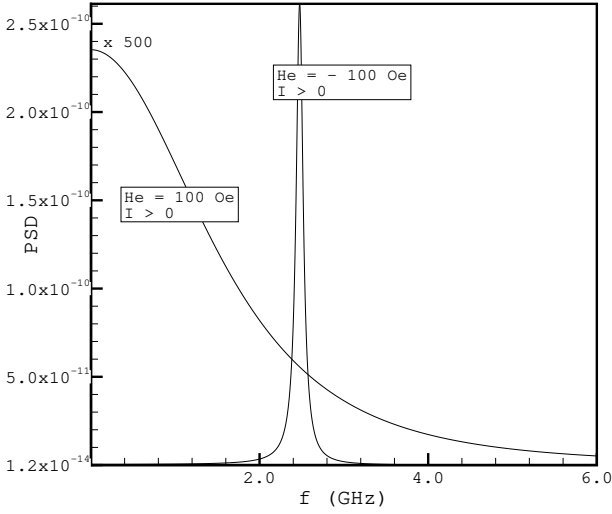,height=8 cm}}
  \end{center}
  \caption{{{Comparison of the PSD (arbitrary units) for positive and negative biasing fields $H_e$.
  \ For $H_e = 100 \; \mathrm{Oe}$  the peak (which is multiplied by 
$500$)
has been completely suppressed and there is a shift to the
left, a signature of an over-damped state. }}}
\label{100n100}
\end{figure}

\begin{figure}[ht!]
  \begin{center}
 \mbox{\epsfig{file=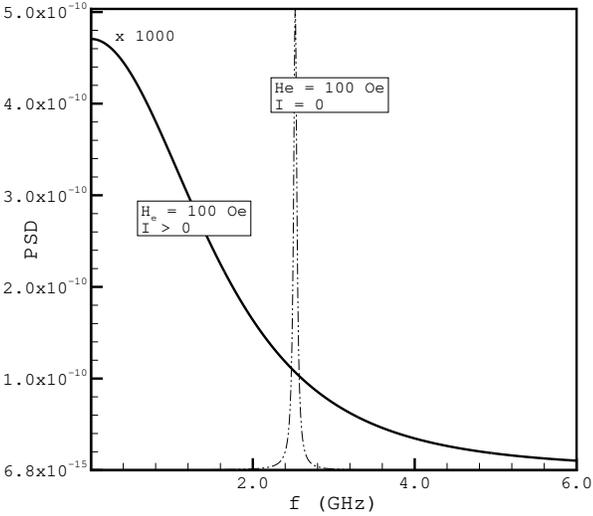,height=8 cm}}
  \end{center}
  \caption{{{ Effect of the spin torque on the PSD (arbitrary units) curve for
$H_e = 100 \; \mathrm{Oe}$. \ The spin torque is clearly acting as a damping force for positive current.
\ For $I=0$, the PSD curve has the familiar FMR shape. }}}
\label{100i0i}
\label{100i0}
\end{figure}

\ In figure \ref{100n100}, we plot the noise for $H_e = 100 \; \mathrm{Oe}$ and $H_e = - 100 \; \mathrm{Oe}$.
Clearly for the case with the positive field, the noise is completely
suppressed
  compared to the case with negative easy axis
biasing. \ This is consistent
with the experiment. \ Therefore the state with positive biasing is
equivalent
to a state with large effective damping. \ This large damping is coming
from the spin momentum transfer. \ If we turn off the current, we get back the
usual FMR (bright) spectrum (see Fig. \ref{100i0}) in this case too. \ This
asymmetry between positive and negative biasing fields close
to the perpendicular direction of the free layer will be important
later when we have a DW in the presence of  spin torques.

\begin{figure}[ht!]
  \begin{center}
\begin{tabular}[c]{ll}
a & \mbox{\epsfig{file=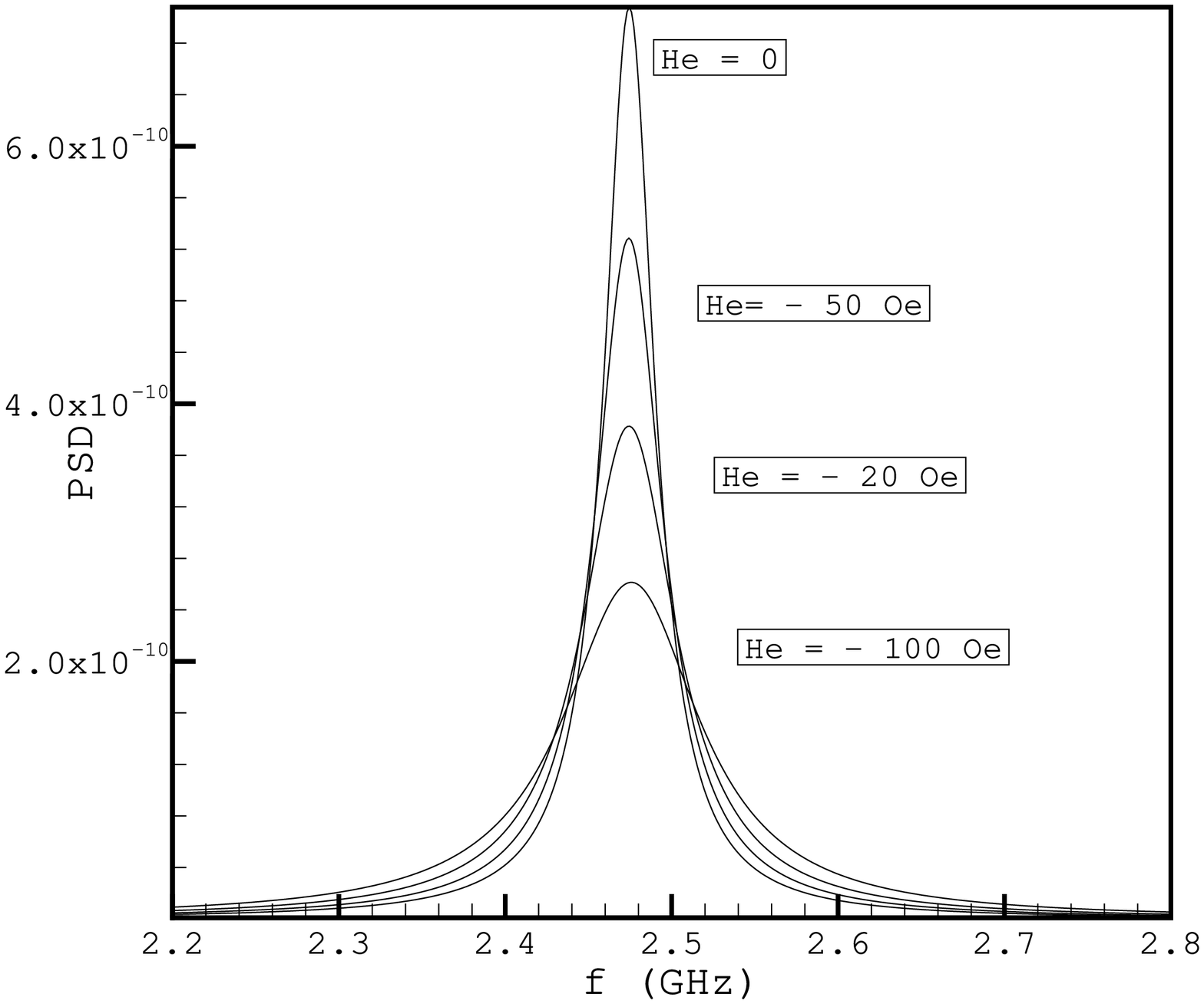,height=8.5 cm}} \\
 b & \mbox{\epsfig{file=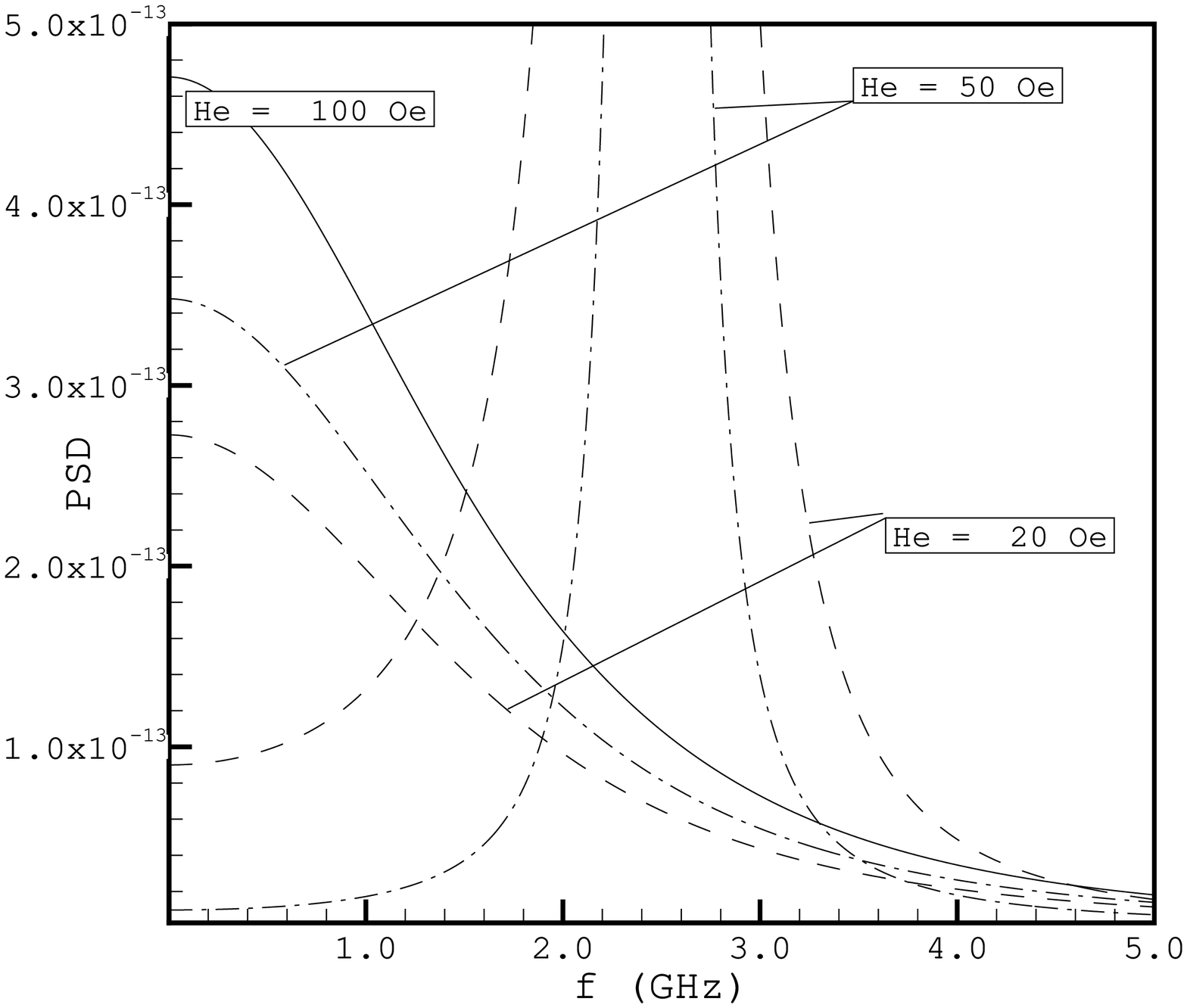,height=8.5 cm}}
\end{tabular}
  \end{center}
  \caption{{{PSD (arbitrary units) for both negative and positive biasing: \ (a) PSD
 (FMR) for negative easy axis biasing
fields. \ (b) PSD curves for positive bias fields.
\ For fields approximately between $50 \; \mathrm{Oe}$ and $20 \;
 \mathrm{Oe}$, there are two possible states for the system; one is over-damped (dark) and the other is regular (light).
Outside this range of fields, only dark or light states exist. }}}
\label{negative}
\label{positive}
\end{figure}

\ Therefore the single spin model captures the 'bright' and 'dark' regions
of the spectral density for frequencies
around the FMR frequency (see Fig. 2 in Ref. \onlinecite{covington}). \ Fig.
 \ref{negative}a shows the strength
of the power as a function of the negative bias field. \ Clearly
for large biasing we have less noise as expected. \ Now, if we plot the same
curve for positive fields, Fig. \ref{positive}b, we find a very
interesting result. \ For large positive fields, we have the usual
'dark' regions that reflect
 high damping states. \ As we lower the field, we find that
the system now can sustain {\it two} states, one bright and one is
dark. The dark state is actually
less stable than the bright one in this
case. \ The x-component of the magnetization in the
dark state is negative, i.e., {\it opposite} to the direction of the easy axis
field while the bright one is along the field $H_e$.  \ This
is most probably the origin of the $1/f$-type noise
in the system. \ The $1/f$ region in the experiment appears on the
negative side of the easy axis field. \ Here it appears on the positive
side \cite{covington}. \ The reason is that the zero point of the
axis is not
well known in the experiment. \ The experiment estimates that the
magnetization is perpendicular to the pinned layer
at $H_e = 34 \; \mathrm{Oe}$ which should correspond to
$-20 \; \mathrm{Oe}$ in
our case. \ Therefore there is a shift of about $ 50 \; \mathrm{Oe}$ in the
reference point which is approximately the field when two
states become  possible as a solution to our equations. \ The
important point we need to remember that the spin
system behaves differently for positive and negative bias
when there is a  spin torque. \ This is mainly due to the fact
that in one case the spin torque is acting as a regular field, while
in the other, it is acting as an extra source of damping.

\subsection{Tri-layer CPP structure with a trapped DW}

\label{subsec:tDW}

Next, we turn to the study of the DW case. First, we show how a DW
in CPP-SV can be manipulated by low currents through the
spin torque. The interaction of the DW with an external field will
be also shown.
\subsubsection{The effect of spin polarized current}

\begin{figure}[ht!]
  \begin{center}
\mbox{\epsfig{file=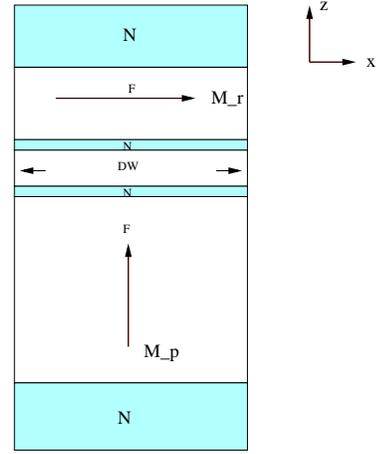,height=6 cm}}
  \end{center}
  \caption{(Online color) Schematic geometry layout for  the 3 magnetic -layers CPP structure.
  The first layer is now polarized
along the current direction. The middle has the DW. The third
layer has been introduced to  enable the  GMR sensing of the DW
motion.} \label{fig222}
\end{figure}

\begin{figure}[ht!]
  \begin{center}
\begin{tabular}[c]{ll}
a &  \mbox{\epsfig{file=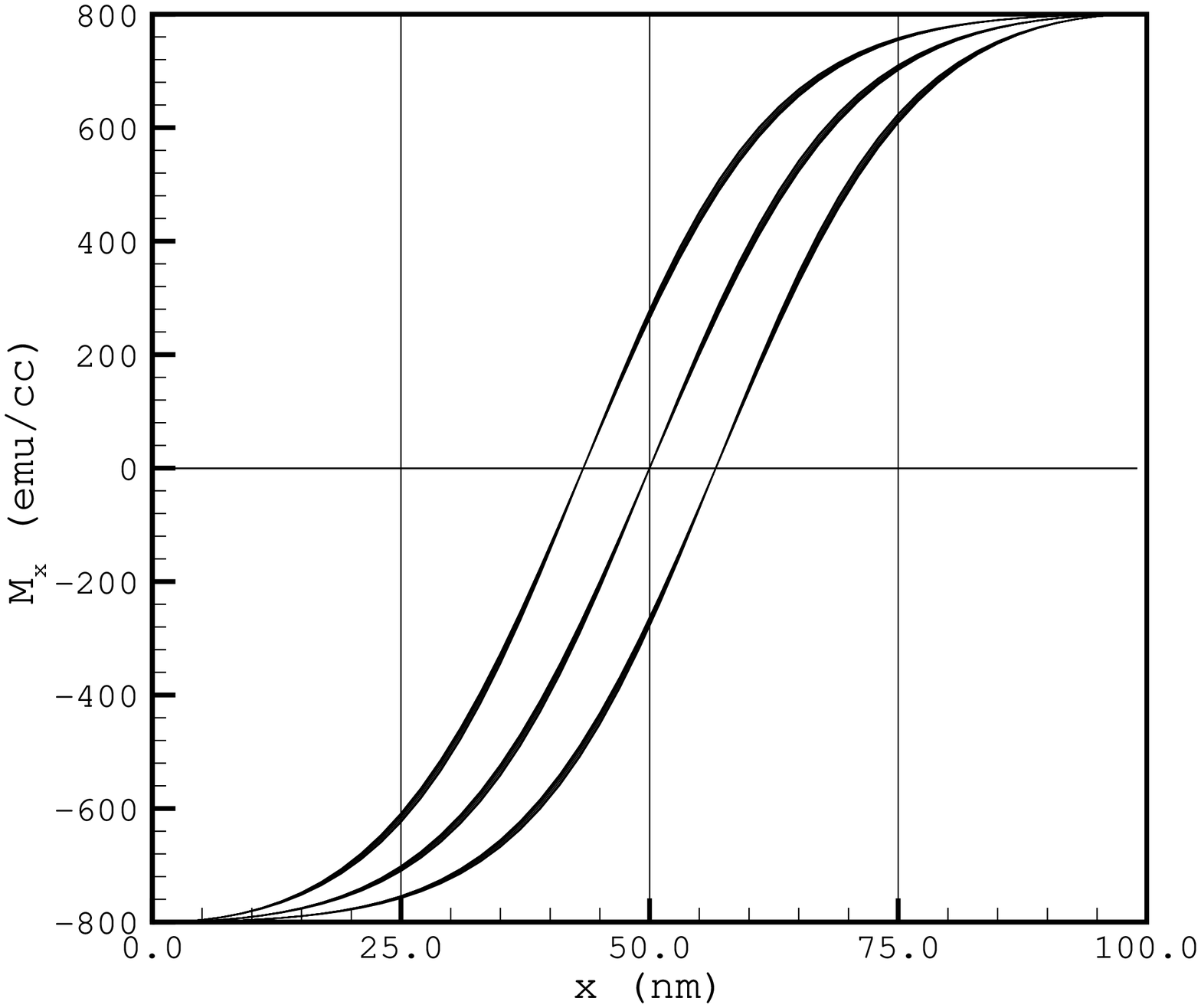,height=8 cm}}\\
b &  \mbox{\epsfig{file=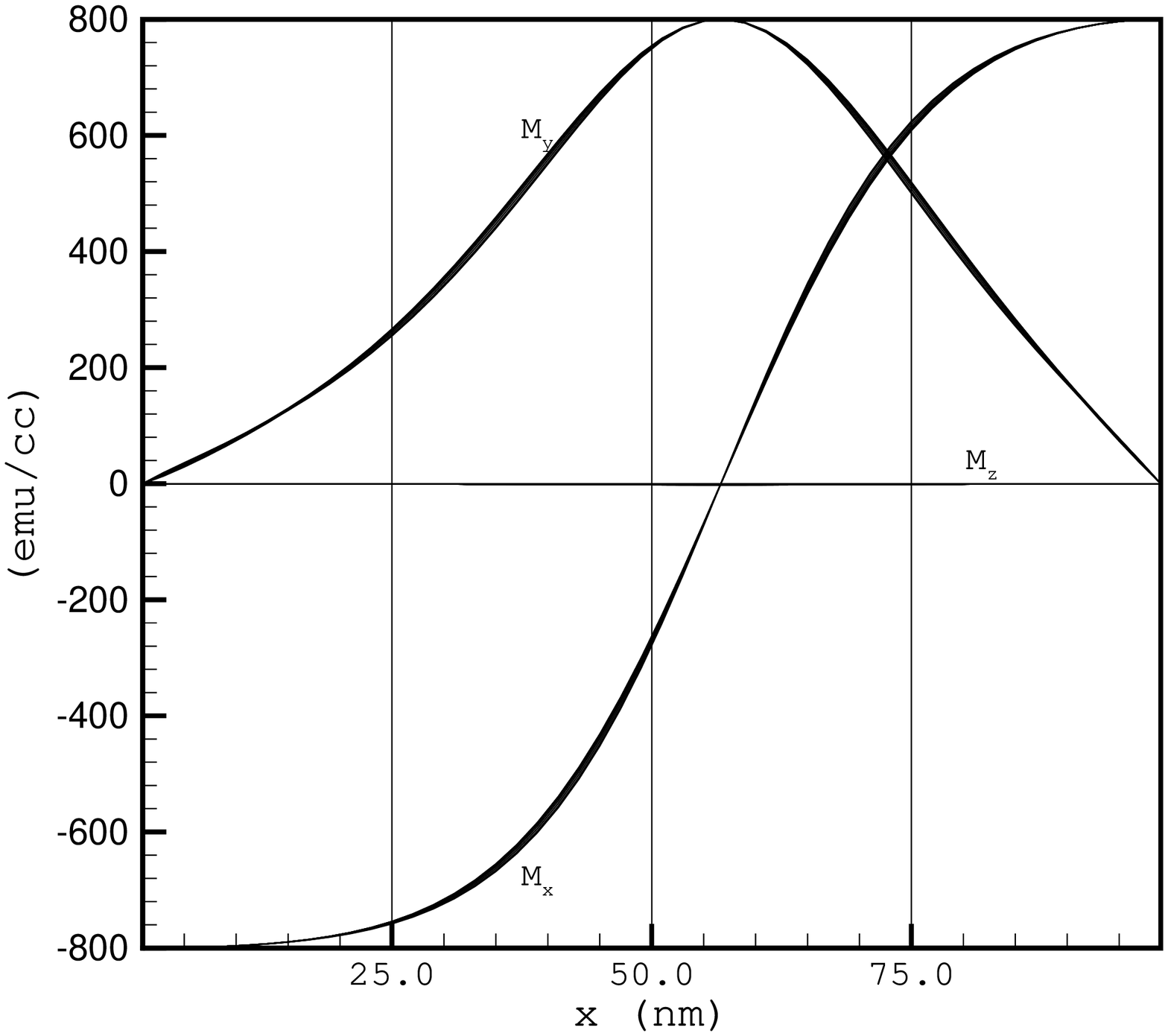,height=8 cm}}
\end{tabular}
  \end{center}
  \caption{Magnetization profiles of the in-plane components as a
  function of the sign of the current in the CPP device. \ In (a), depending 
on the sign of the current the curve shifts to the left or to the right. b) The
 shape of the x and y components of the magnetization in the DW is a slight 
distortion from that without currents.
 }
\label{fig2}
\end{figure}

\begin{figure}[ht!]
  \begin{center}
 \mbox{\epsfig{file=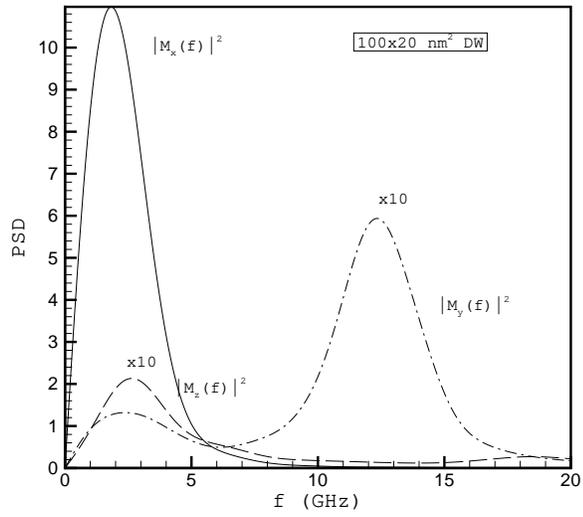,height=8 cm}}
\end{center}
\caption{PSD (arbitrary units) of the different components of the magnetization in
the presence of current $I=0.001 \; \mathrm{mA}$ for a constrained DW.
} \label{noisefig}
\end{figure}

First, we consider  an alternative CPP structure. As will be shown
below this modification of the traditional CPP structure can be
done for at least three reasons. One reason is to create structures
where  DW can be manipulated with spin momentum in most efficient
way. Secondly, we would like to be able to detect domain wall
motion with GMR effect. We also find that the suggested CPP structure
modifications may have some advantages in terms of reducing
effects of magnetization instabilities due to the spin momentum
transfer.
\ As has been shown, for example in Ref. \onlinecite{rebei2},
even in CPP devices with nominally uniform magnetization, the spin
torque can give rise to magnetization instabilities. Thus, the
latter reason should be kept in mind as an important one.

\ In the following we consider a CPP structure which has three
magnetic layers (Fig. \ref{fig222}) where  the DW layer is
sandwiched between two pinned magnetic layers with one of them
polarized along the direction of the current and the other
polarized along the easy axis of the middle DW layer.
\ In the following simulations, the bottom layer is taken to be
$100 \times 20\times 9 \; \mathrm{nm}^3$, the middle layer is
$100\times\times 20\times 2 \; \mathrm{nm}^3$, and the reference layer has
the dimensions $100\times 20 \times 3 \; \mathrm{nm}^3$. \ In this
geometry, the two outer magnetic layers lead to a  two different
spin torques, $\Gamma = \Gamma_B + \Gamma_T$, acting on the middle
magnetic layer
\begin{equation}
 \mathbf{\Gamma} = - p I \left( \mathbf{m} \times \left( 
\mathbf{m}\times 
\mathbf{m}_B \right) -
\mathbf{m} \times \left( \mathbf{m}\times \mathbf{m}_T \right) \right),
\end{equation}
where $\mathbf{m}_B$ ( $\mathbf{m}_T$) is the magnetization direction
of the bottom
(top) layer. \ The damping parameter in this section has been
increased to $\alpha = 0.08$ to better account for spin
accumulation \cite{rebei}.


\ First, we investigate if  the DW can be  moved along the easy
x-axis with  moderate  currents.
\ This will enable  a spin torque $\Gamma_B=-\mathbf{m}\times
\mathbf{h}_{sp}^B$ with an effective field along the x-axis and
proportional to the y-component of the magnetization in the DW
layer that is largest at the center, $\mathbf{h}_{\mathrm{sp}}^B = p I
( m_y,- m_x, 0 )$. \ Since $M_y \approx M_s$ around $x=0$, this gives
us the optimal field needed to push the DW off the center and this
appears to be a primary  reason why only
 very low  currents are needed
to have an appreciable motion of the DW in considered CPP
geometry.
\ Figure \ref{fig2} shows the effect of the spin torque on the DW
in the 3-layer geometry as a function of the current. \ We find
that the spin torque from the top layer has a
 relatively small
effect on the dynamics of the DW since its effective field is
$\mathbf{h}_{\mathrm{sp}}^T = p I (0,- m_z, m_y)$. \ Given that the
z-component of the magnetization is practically zero for the
currents in the case shown in Fig. \ref{fig2}, the effect of
$\mathbf{h}_{\mathrm{sp}}^T$ on the magnetization is negligible. \ We find
that indeed in this geometry, the spin torque can be used to
control the motion of  DW with very small current densities. \
This is primarily  due to the fact that the constrained DW has
a non-zero 
y-component of magnetization in the DW region.

\ The displacement of the DW by the spin torque (we make sure that
the  Oersted field is not the origin of this motion) is easily
understood from the equation of motion without the demagnetization
field. \ Taking account of only
  the spin
torque, the exchange and the anisotropy, the static equations for
the magnetization are
\begin{eqnarray}
\frac{2 A}{M_s^2} \frac{d^2 m_x}{d x^2}-\frac{2 K}{M_s^2} m_x + p I m_y
 & = & c m_x ,\\
\frac{2 A}{M_s^2} \frac{d^2 m_y}{d x^2}- p I  (m_x - m_z)  & = & c m_y ,\\
\frac{2 A}{M_s^2} \frac{d^2 m_z}{d x^2}+ p I  m_y  & = & c m_z,
\end{eqnarray}
where $c(x)$ is a real function and $ \mathbf{m}^2 = 1$. \
Clearly, in this case the x-component is coupled to the y
component which acts as a source term for the x-component. \
Neglecting anisotropy and integrating the equation for the $m_x$
component around zero, we find that the difference in the slope of
$m_x$(x) for x=0 and small $x=\epsilon $  is given by
\begin{equation}
\frac{d m_x}{d x}|_{x=0} - \frac{d m_x}{d x}|_{x=\epsilon} = p I
\epsilon m_y .
\end{equation}
 For positive $\epsilon$, i.e. a shift to the right, the slope at
$x = 0$ is smaller than that at $x=\epsilon$ which is
approximately equal to that at $x=0$ in the absence of  current. \
Therefore $I$ should be negative for positive $m_y$ which is
approximately equal to 
the one around $x=0$. \ This is confirmed by the numerical
integration of LL equation in fig. ~\ref{fig2}.
 The spin torque can therefore be used
 to move the DW in a controlled fashion with low currents. \
 In addition we find, that a three-layer structure may actually have
 lower frequency noise in the 
presence of spin torque than the structure investigated in Ref.
 \onlinecite{rebei2}. \ This potential advantage of our 
proposed  structure
is however realized only  if the  middle layer geometrical
dimensions are comparable with  the DW width, Fig. \ref{noisefig}. \
In this case the PSD in the x-component does not show any
substantial low frequency noise. For the parameters used here, we find that 
the DW width is approximately $40$ nm. The dimension of the film 
is $100$ nm. Therefore the DW is barely constrained and hence the reason behind
the sensitivity of the DW to external forces due to fields or currents.

\clearpage

\begin{figure}[ht!]
  \begin{center}
\begin{tabular}[c]{llll}
a &  \mbox{\epsfig{file=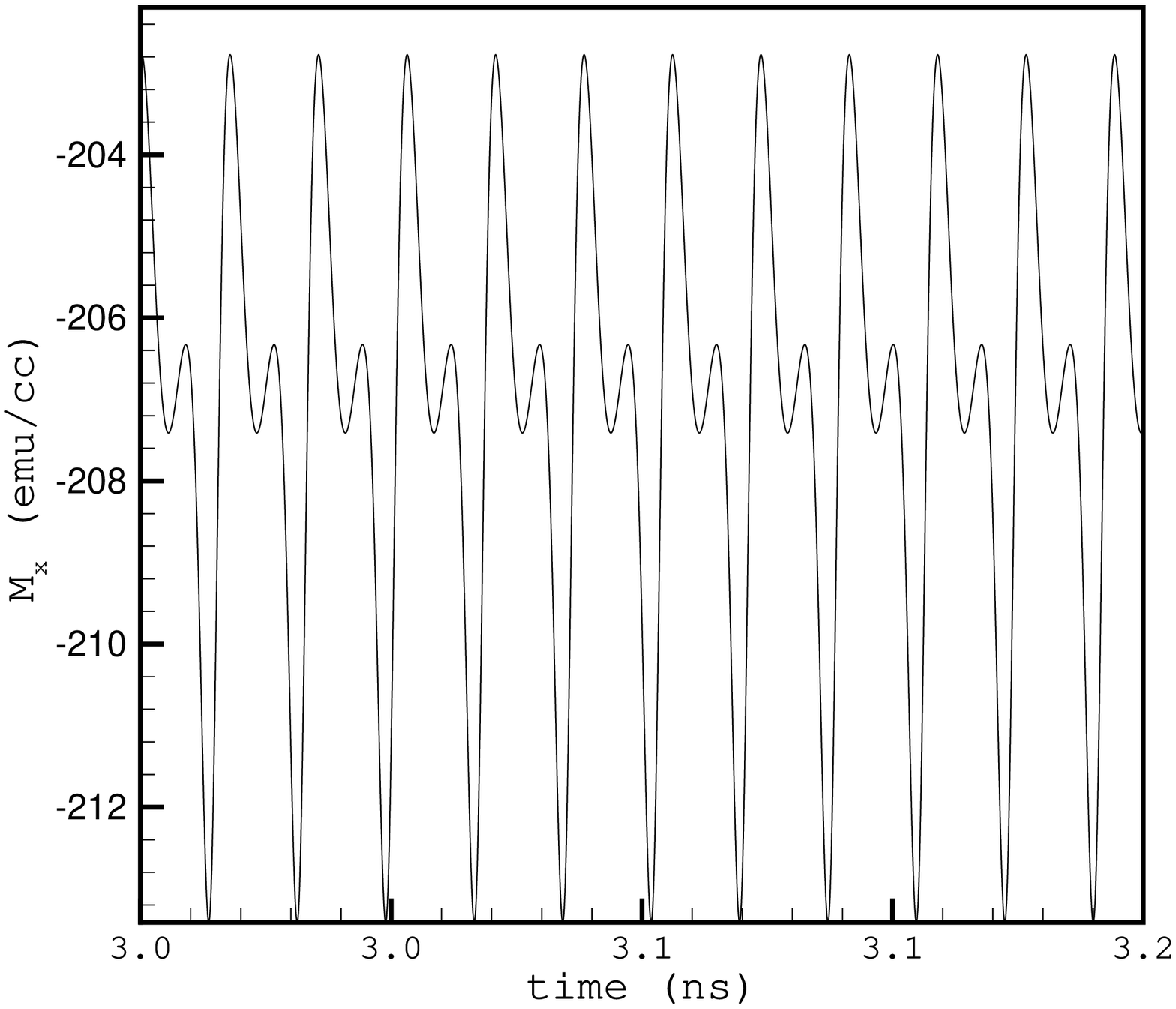,height=7.5 cm}} &
b & \mbox{\epsfig{file=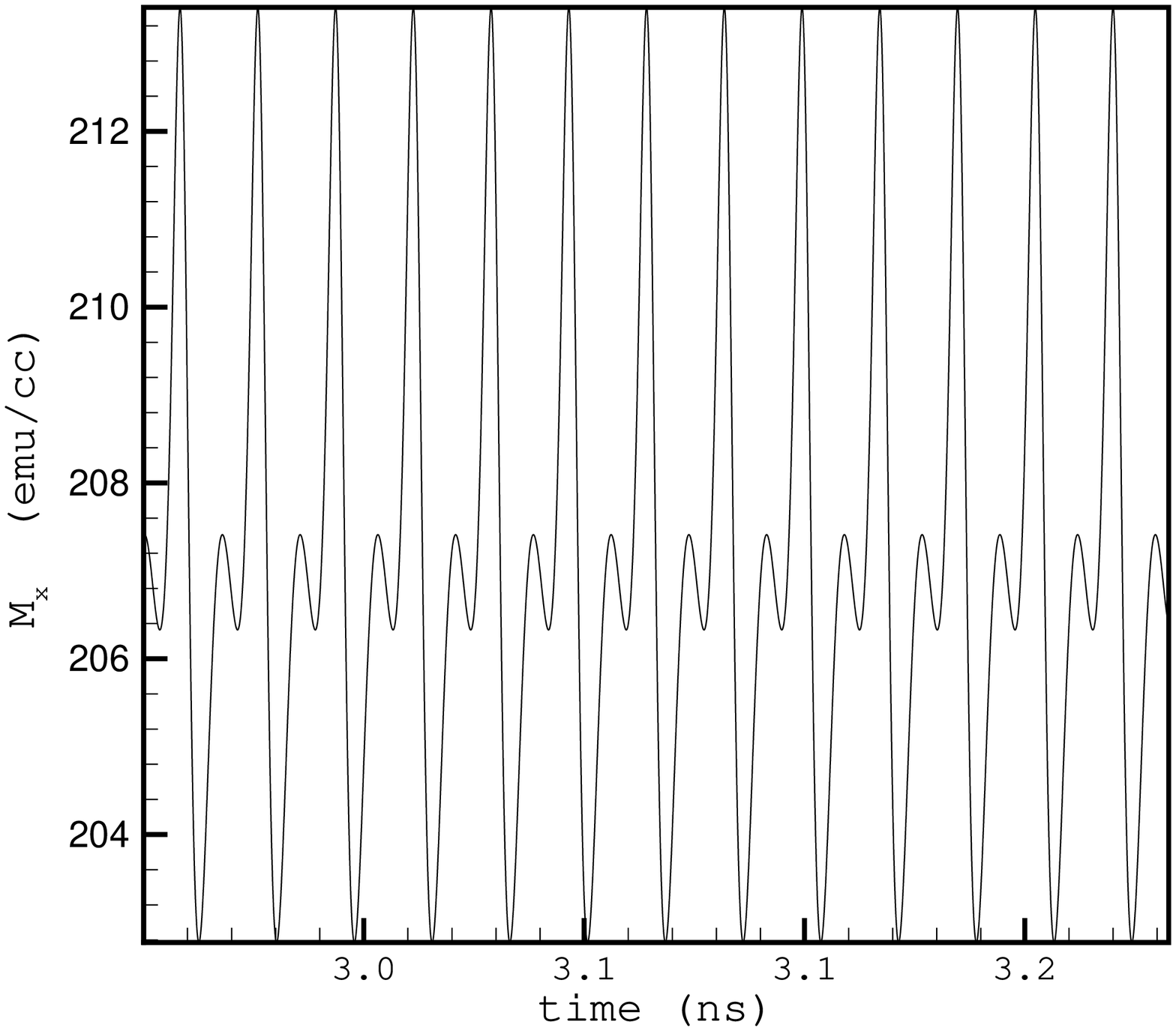,height=7.5 cm}}\\
c & \mbox{\epsfig{file=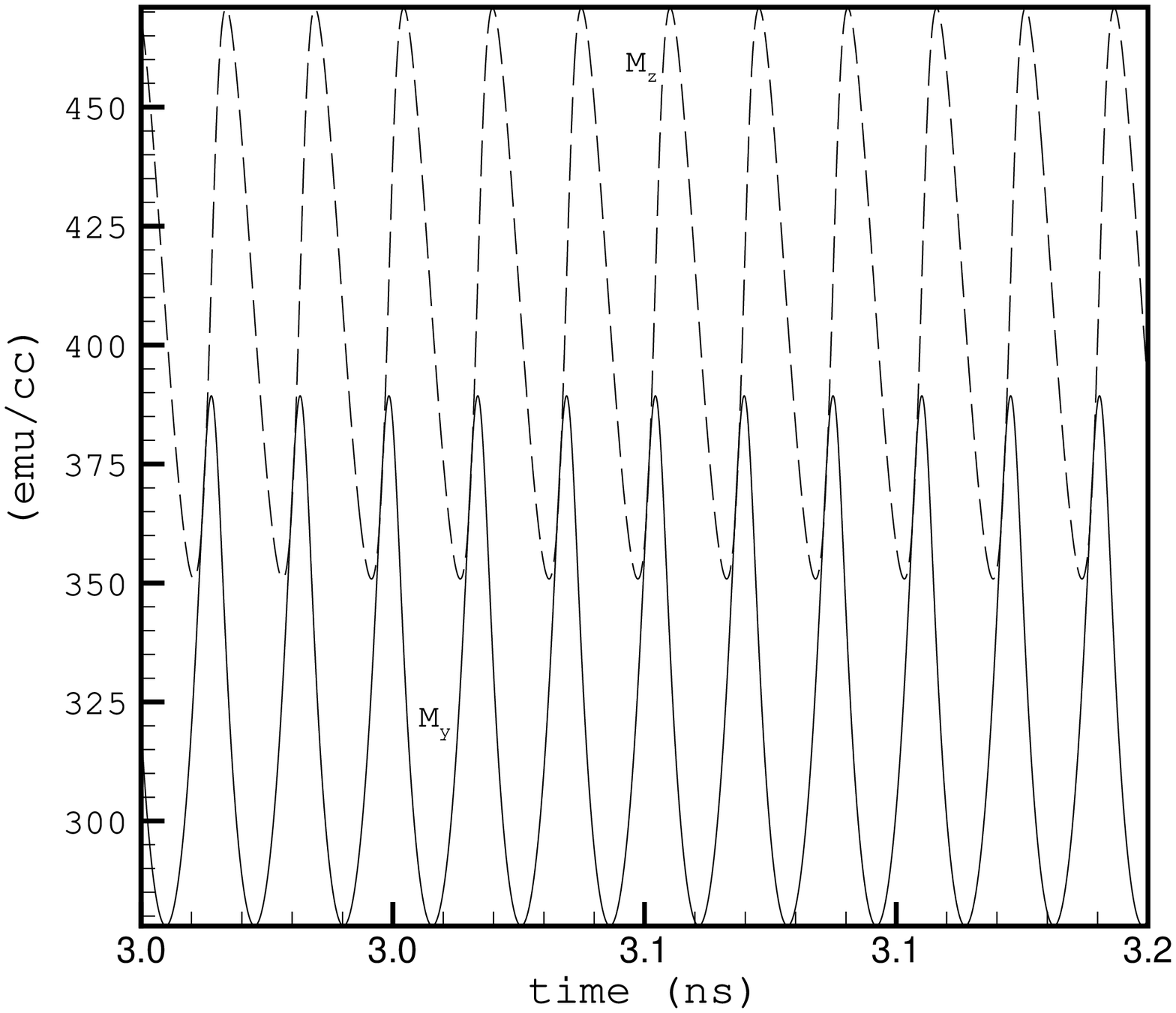,height=7.5 cm}} &
d & \mbox{\epsfig{file=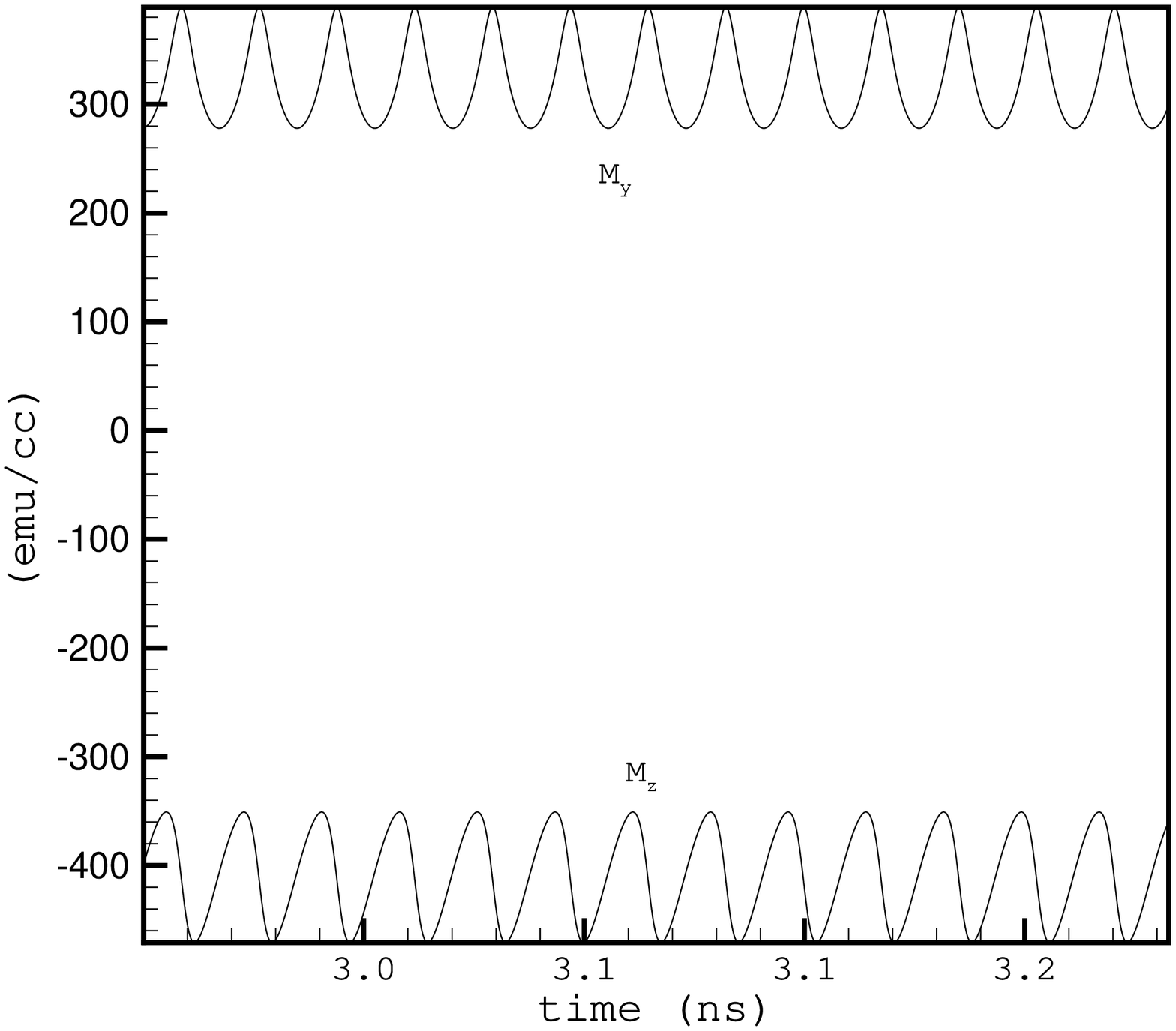,height=7.5 cm}}
\end{tabular}
\end{center}
  \caption{The time evolution of the different components of the average magnetization of the
  DW for $I = 0.1 \;  \mathrm{mA}.$ (a,c) and $ I = -0.1 \; \mathrm{mA}$ 
(b,d).
  No stationary solutions exist at high currents.}
\label{highI}
\end{figure}

\clearpage

\ At higher currents, we are no longer in a linear regime. \  The
numerical integration of the LLG equation shows that the z-component
of  magnetization becomes more significant as we increase the
current and this contributes to the twist of the DW and no
stationary solutions are possible in this case. \ Figure
\ref{highI}  shows the time evolution of the magnetization for
current densities of the order of $ 1.0 \times 10^7 \; \mathrm{A}/\mathrm{cm}^2$.

\ The magnetization dynamics is a regular periodic  rotation. \ In this
case the spin torque can be used to selectively excite
higher modes of the magnetization as compared to those studied in
section II.

\begin{figure}[ht!]
  \begin{center}
\begin{tabular}[c]{llll}
a & \mbox{\epsfig{file=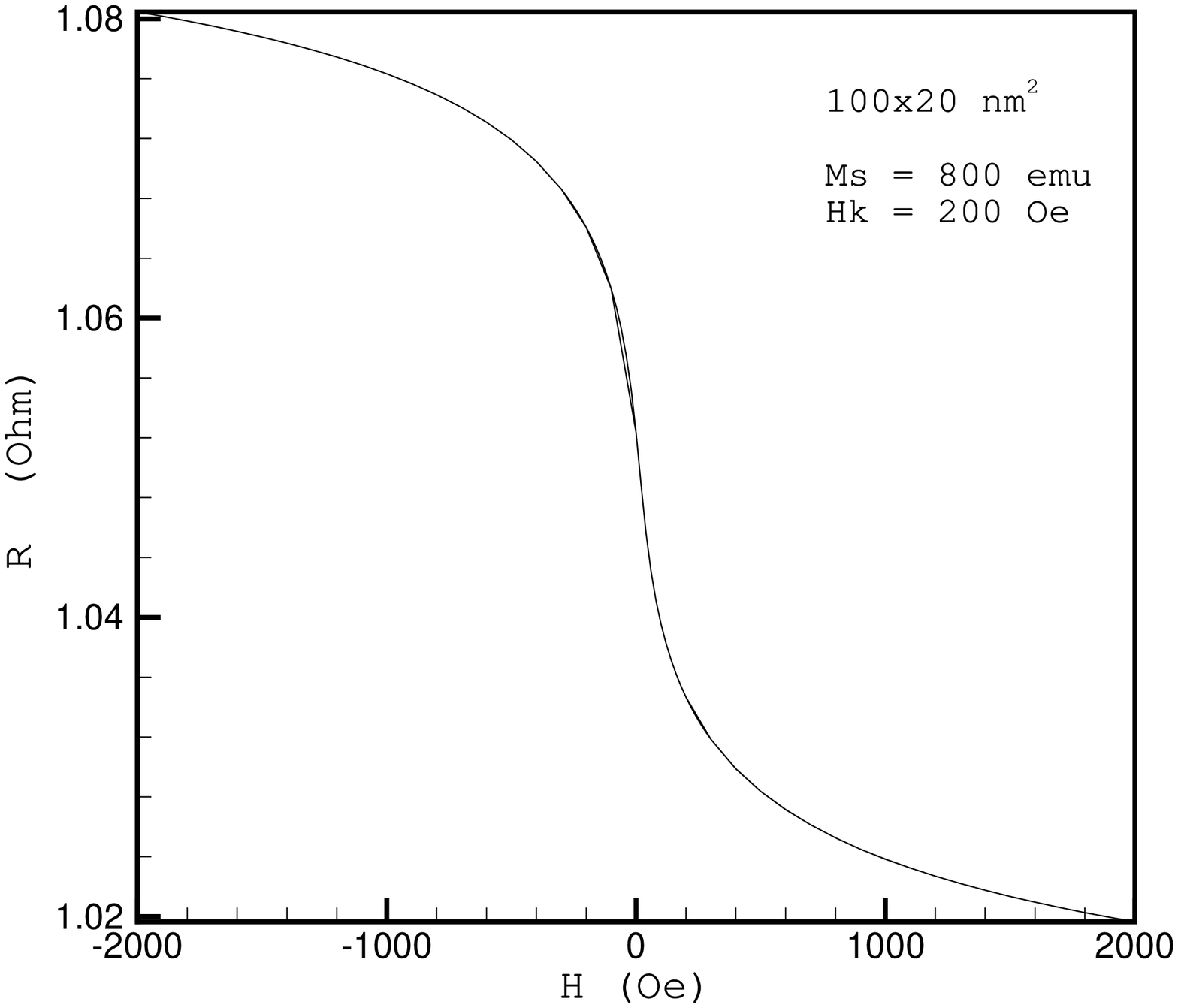,height=8 cm}} &
b & \mbox{\epsfig{file=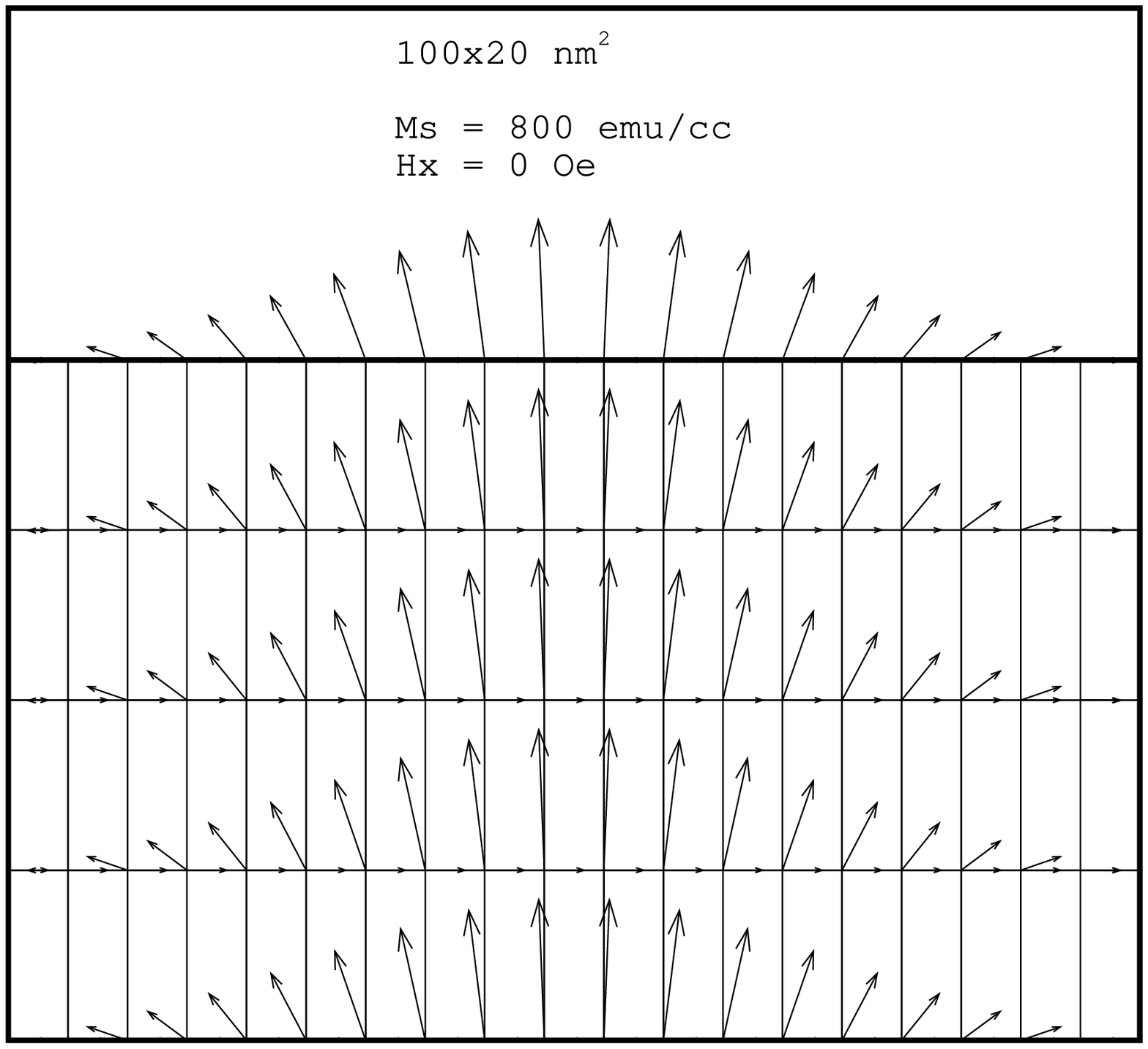,height=8 cm}}\\
c &  \mbox{\epsfig{file=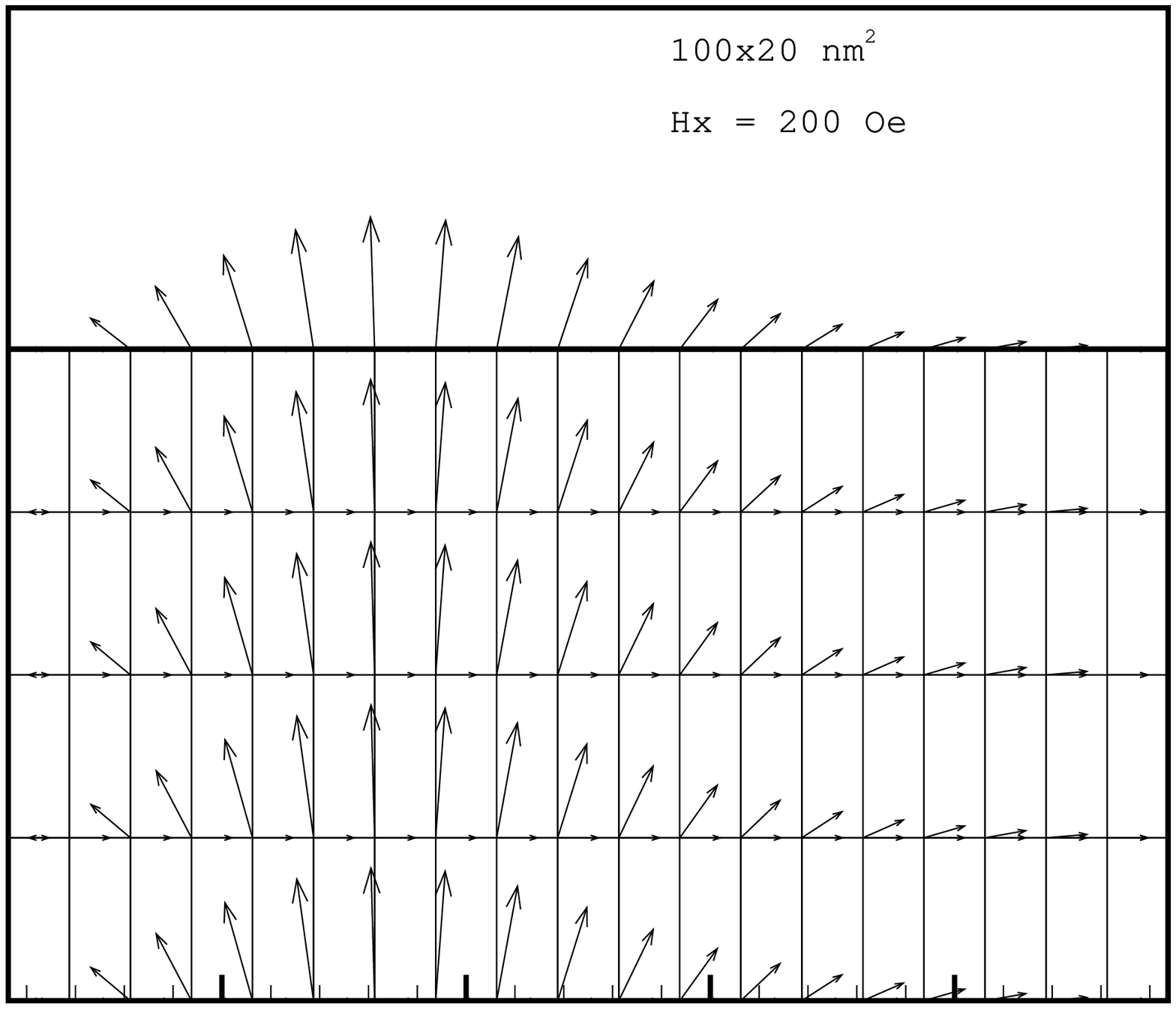,height=8 cm}} &
d &  \mbox{\epsfig{file=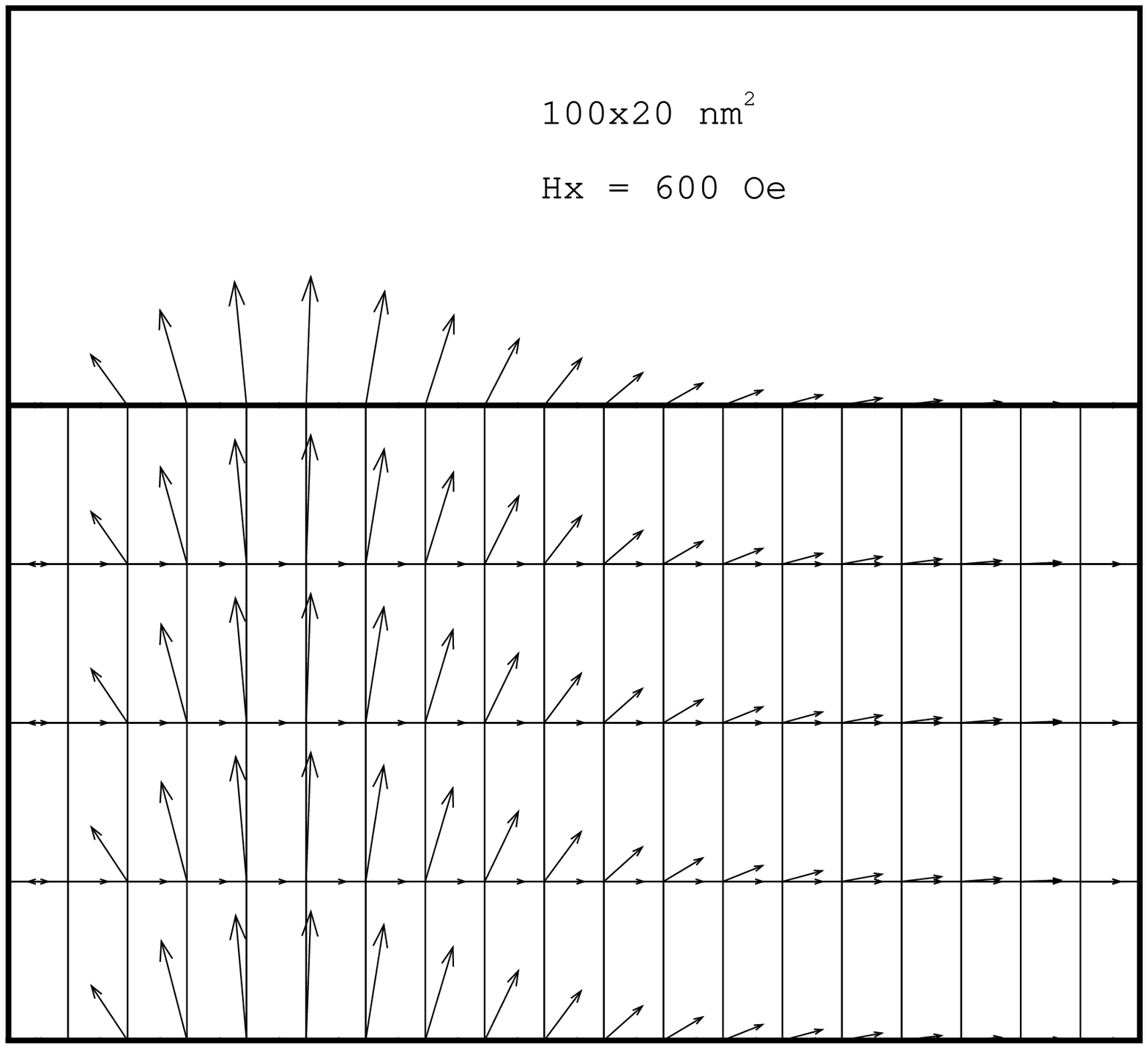,height=8 cm}}
\end{tabular}
  \end{center}
  \caption{{{ (a) Resistance versus external field applied along 
the  easy x axis for the $100\times 20 \; \mathrm{nm}^2$ film. \
  The reference layer is pinned in the $+x$ direction.
  The resistance is normalized in such a way that for $R=1$ the two layers are parallel and for $R=1.1$ they are anti-parallel. \
  (b-d) Profiles of the magnetization in the $xy$ plane 
for  zero (b) and positive 
fields (c-d), respectively. \ The scale for the x and y axis are different. \ For a given x, five points are plotted along the y-axis. \ The DW is fairly uniform along the y axis.   }}}
\label{fig1}
\end{figure}

\subsubsection{The Effect of External Magnetic Field on a DW}

\ Finally in this section, we investigate effect of an external
magnetic field. We add a Zeeman term to the total energy and study
the displacement of the DW due to an external field along the easy
axis.

\  The  external field along the easy axis is applied to  the
middle layer in the presence of a small current to measure
resistance changes across the CPP structure and so that no spin
torque effects are appreciable on the DW.
\ Interestingly the three-layer structure with DW  does not require
biasing which is needed for standard CPP structures to achieve
linear dependence of resistance on the external field.
\ The calculated transfer curve of resistance $R$ versus field $H$
is shown in Fig. \ref{fig1}. As can be seen this dependence  is
centered around zero, has a large slope $d R/ d H$, and shows
small  hysteresis. \ Moreover, the system appears to be more stable
to perturbations by the spin torque and no 1/f-type behavior is
observed in this case. Our device is therefore well suited to 
function as a magnetic sensor. However the proposed structure 
lacks an important property which is
 needed in memory 
applications and that is non-volatility. As we 
remove the voltage 
across the CPP-SV, the DW relaxes back to its 
equilibrium position and hence any state stored in the 
DW  position   
is lost. Nevertheless, our proposed CPP-SV structure
 can be incorporated as 
part of a logic device. Recently, properly redesigned CPP 
structures have been proposed for reprogammable 
logic 
elements \cite{ney,moodera}. This latter  application 
does not require non-volatility and hence 
our device can be utilized in a similar way as 
in Refs. \onlinecite{ney,moodera}. Our device  has an advantage 
compared to that proposed in Ref. \onlinecite{ney} and that is only 
much smaller currents are needed in our case.

\clearpage

\clearpage

\section{Summary and  Discussion}

\label{sec:summ-conc}

\ In summary, we have presented a study of magnetization dynamics
for CPP geometry which includes a constrained DW layer.
\ We have identified a Doring-type mode and a new breathing mode.
\ It is shown that the lowest modes of the DW dynamics  can be
understood in terms of the parity of the inhomogeneous 
ground state.\ We have
investigated in details how the constrained DW dynamics is affected
by the spin polarized current and thermal fluctuations and compared it
with the traditional single domain free layer structures.
 \ In
particular, we found that the currents needed to measure any
appreciable motion of the DW are at least two orders of
magnitude less than usual values of currents needed to switch the
single domain magnetization. \ This difference is attributed to
the appearance of a significant
magnetization component of the constrained DW that  is
perpendicular to the pinned layer magnetization and the exchange field.

\ We also find that  thermally activated  motion of the constricted
DW has lower weight in the lower frequency  region than that of
the unconstrained DW. The latter shows well known telegraph type
noise characteristics.  This difference can be understood using
notion of charged DW introduced by Neel \cite{Neel_dw} and
competition of the magnetostatic and size dependent exchange
interaction contributions to the DW   potential for motion along
the easy axis (see Fig. ~\ref{demag}) .
 The  three magnetic layer CPP structure was
 introduced  so that
the spin torque effect on  DW layer  is maximized .  \ This CPP
geometry has been investigated and found to have a number of
interesting properties such as (i)
DW can be easily controlled by an external field or a polarized
current with relatively small current densities; (ii) linear
dependence of resistance on the external  field and current; (iii)
improved magnetization stability characteristics. Experimental 
realization of the device proposed here
 requires finding 
ways to constrain the DW within the middle 'free' layer. The pinning 
at the edges can be realized by creating permanent magnets with 
different coercivity and/or with 
 anti-ferromagnetic (AF) coupling. The 
AF coupling at the edges needs antiferromagnets with different 
  Neel temperatures
 \cite{ambrose}
 so that a properly designed field-cooling procedure 
could lead to the pinning
in opposite directions at the edges of the magnetic 
stripe.
Other alternatives such as special shaping and padding also 
have been discussed in the 
literature in the context of stability of DW in magnetic 
nano-elements \cite{klaui2}.

\bigskip

\ We thank G. Parker for making his LLG solver available
to us. \  P. Asselin and W. Scholz have provided us with
 many helpful comments on the text.  \  We also acknowledge 
helpful discussions
with T. Ambrose, L. Berger and
M. Covington.

\end{document}